\documentstyle[prd,aps,floats,epsfig]{revtex}  
\def\DESepsf(#1 width #2){\epsfxsize=#2 \epsfbox{#1}}  
\def\msq{\widetilde m}
\def\xgq{x_{\tilde g\tilde q}}
\def\mgl{m_{\tilde g}}
\begin{document}  
%
\draft  
\twocolumn[\hsize\textwidth\columnwidth\hsize\csname
@twocolumnfalse\endcsname
\title{One-Loop Analysis of FCNC 
       Induced by Right-handed Down Squark Mixings}  
\author{Chun-Khiang Chua and Wei-Shu Hou}  
\address{  
Department of Physics , National Taiwan University,   
Taipei, Taiwan 10764, R.O.C.  
}  
\date{\today}  
\maketitle  
\begin{abstract}  
If the underlying flavor symmetry is Abelian, 
quark mixings in $d_R$ sector are the most prominent.
Such flavor violating effects
can reveal itself through $\tilde d_R$ squark mixings if 
supersymmetry is realized in Nature. 
Quark-squark alignment is necessary to deal with $\Delta m_K$ 
and $\varepsilon_K$ constraints, but interestingly, 
with $m_{\tilde q}$, $m_{\tilde g} \sim$ TeV, 
the $\tilde d_R$ mixing effects are 
comparable to $B_d$ and  $B_s$ mixings in the Standard Model, 
while $D^0$ mixing is tantalizingly close to some hints from data.
$CP$ phases in these mixings would therefore be deviant, and 
$\vert V_{td}\vert$ and $\arg V_{ub}^*$ may be 
larger than allowed by unitarity constraints, 
which can be checked by the BaBar and Belle experiments.
Mixing induced 
$CP$ violation in $b\to s\gamma$ and $d\gamma$ transitions
can be obtained,
in particular, by sizable enhancement with an extra $\tan\beta$ factor from
non-standard soft breaking terms.   
Heavy superparticles can escape present flavor changing neutral
current (FCNC) bounds and direct searches at colliders, 
but reveal themselves in the B system. 
\end{abstract}  
\pacs{PACS numbers:   
12.60.Jv, 
11.30.Hv, 
11.30.Er, 
13.25.Hw  
}]   
%
%
%
%
%
%
\section{Introduction}    
Despite its great success, the Standard Model (SM)
is widely regarded as a weak scale effective theory.
We expect to encounter new phenomena that arise from a more complete theory
as we increase the luminosity and energy of our probes.
Although we have not observed any convincing deviation of experimental results
from SM predictions (except neutrino oscillations \cite{superK}),
there are a few hints on the existence of New Physics.   

It is well known that the sine of the $CP$ violating phase,
$\phi_1$ (or $\beta$) $\equiv \arg V_{td}$ (PDG phase convention~\cite{PDG}), 
can be measured via the time dependent asymmetry,  
\begin{eqnarray}  
a_{J/\psi K^0_S}&=&{\Gamma (\bar B^0(t)\rightarrow J/\psi K^0_S)-  
                  \Gamma (B^0(t)\rightarrow J/\psi K^0_S) \over  
                        \Gamma (\bar B^0(t)\rightarrow J/\psi K^0_S)+  
                        \Gamma (B^0(t)\rightarrow J/\psi K^0_S)}
\nonumber \\  
               &=&\sin 2\phi_1\, \sin \Delta m_{B_d} t.  
\end{eqnarray}  
Following earlier measurements \cite{CDF,OPAL,ALEPH,Belle,BaBar},
the BaBar and Belle Collaborations have recently 
firmly established \cite{BaBar2,Belle2} $\sin 2\phi_1$ to be nonzero.
Compared to earlier low results of $\sin 2 \phi_1=
0.58{+0.32+0.09\atop-0.34-0.10}$ (Belle) \cite{Belle} 
and $0.34\pm0.20\pm0.05$ (BaBar) \cite{BaBar},
the corresponding numbers are now $0.99\pm0.14\pm0.06$ (Belle) \cite{Belle2} 
and $0.59\pm0.14\pm0.05$ (BaBar) \cite{BaBar2}, respectively.
Combining the two most recent values without systematic errors,
one gets the average value
\begin{equation}
\sin 2 \phi_1=0.79{\pm 0.10}.
\label{sin2phi1}
\end{equation}  
While this is still consistent with 
the Cabibbo--Kobayashi--Maskawa unitarity (CKM) fit value of 
$\sin 2\phi_1=0.698\pm0.066$ \cite{Ciuchini:2000de}, 
the central value is now somewhat on the high side, 
especially for the Belle number.  
If this trend persists --- which we would know by summer 2002 ---  
it would imply the presence of New Physics. 
With this in mind, it is clearly a good time to  
study other related $CP$ violation processes,  
especially those that are suppressed in the SM. 

It has recently been shown that charmless rare $B$ decays favor \cite{Hou} 
a value for $\phi_3$ (or $\gamma$) $\equiv \arg V_{ub}^*$ 
that is larger than the one obtained from
the CKM unitarity fit \cite{Ciuchini:2000de}.  
The latter is dominated by recent improved   
bounds on $\Delta m_{B_s}/\Delta m_{B_d}$ \cite{LEPBOSC},
\begin{eqnarray}
\Delta m_{B_d}&=&0.484\pm 0.010\,{\rm ps}^{-1},
\\
\Delta m_{B_s}&>&15.0\,{\rm ps}^{-1}\,{\rm at\,\,95\%\,\,C.L.},
\end{eqnarray}
which tends to squeeze out the large $\phi_3$ possibility.
One also has Br$(K^+\to\pi\nu\bar\nu)=4.2^{+9.7}_{-3.5}\times 10^{-10}$
from the E787 Collaboration \cite{E787}, where
the central value is several times above the SM expectation,  
implying a rather large $|V_{td}|$.
The last branching ratio, of course can be viewed as decreasing 
with time since no new events have been found.
Combining the above, however, 
perhaps a more consistent picture would be if 
$B_d$ or $B_s$ mixings have additional New Physics sources.  
This may already be indicated by the measurement of $\sin2\phi_1$
in Eq. (\ref{sin2phi1}) as we have discussed.
It can alternatively be tested in the $CP$ phase of $B_s$ mixing,
which can be studied at the Tevatron collider in the next few years.
If a non-vanishing value is found,
it would definitely imply New Physics since the SM prediction is zero.
  
The E791, CLEO, FOCUS, Belle and BaBar Collaborations have reported
search results for $CP$ asymmetries in the neutral $D$-meson system
\cite{E791,xD1,xD2,CLEOycp,Belleycp,BaBarycp}. 
The search for $D^0$--$\bar D^0$ mixing using CLEO II.V data suggests
that $x_D\equiv \Delta m_D/\Gamma$ is less then 2.9\% at 95\% C.L.
\cite{xD1}, which is far below previous results \cite{PDG2000}.    
The actual numbers, however, are
$x_D^{\prime2}/2<0.041\%$ and  $-5.8\%<y_D^\prime<1.0\%$,  
in terms of
\begin{eqnarray}  
x_D^\prime&=&x_D\cos\delta_D+y_D\sin\delta_D  
\nonumber \\
y_D^\prime&=&y_D\cos\delta_D-x_D\sin\delta_D,  
\label{xpyp}
\end{eqnarray}  
where $\delta_D$ is the relative strong phase between 
the doubly Cabibbo suppressed $D^0\to K^+\pi^-$ and 
the Cabibbo favored $\bar D^0\to K^+\pi^-$ decay amplitudes.  
The SM predictions of these mixing parameters are small,
$x_D\sim 10^{-5}-10^{-4}$ and $y_D\sim 10^{-4}-10^{-2}$ \cite{smdmix}
(a recent discussion suggests $y_D \sim 1\%$ \cite{Ligeti}).
The CLEO Collaboration arrived at $|x_D|<2.9\%$ by 
assuming $\delta_D$ is as small as suggested by SU(3) and other
arguments  \cite{Wolf_D,Browder}. 
If one removes~\cite{strongphase}   
the prejudice that $D^0 \to K^+\pi^-$ and $K^-\pi^+$   
amplitudes have the same strong phase,
the result on $y_D^\prime$ may actually be hinting at a sizable $x_D$,
which would strongly suggest short distance New Physics interactions.
It is therefore important to compare $x^\prime_D,\,y^\prime_D$ with other
$D^0$--$\bar D^0$ mixing related measurements.

The CP asymmetry parameter $y_{CP}$ is related to the lifetime difference
between $D^0\to K^- \pi^+$ and $D^0\to K^-K^+$,
where the former is flavor specific and the latter a $CP$ eigenstate. 
The FOCUS Collaboration had found the intriguing value of
$y_{CP} =(3.42 \pm 1.39 \pm 0.74)\%$ \cite{xD2}, 
which has a significance of more than $2\sigma$. 
However, recently, the CLEO, Belle and BaBar Collaborations find
much lower values.
The current world average is $(1.1\pm0.9)$\%\cite{Ligeti}.
Still, the point remains that one could
have large $x_D$ if $\sin\delta_D$ is sizable \cite{Bergmann}.

The processes $b\to s\gamma,d\gamma$ occur only at loop level in  
the SM, therefore they are naturally sensitive to New Physics.  
The processes $B\to K^{*}\gamma $ and $b\to s\gamma$  
have long been observed \cite{1st expt}.   
However, the Belle Collaboration observes a $3.5\pm2.1$ 
event excess for $B\to\rho\gamma$, giving 
Br$(B\to\rho^0\gamma)<10.6\times10^{-6}$\cite{Bellerhogamma}
at 90\% C.L.
The quark level $b q \gamma$ coupling is   
usually parametrized as   
\begin{eqnarray}  
H_{\rm SM}&=&-{G_{F}\over\sqrt{2}}  
       {e\over{4\pi^2}} V_{tb}V_{tq}^{*} \,m_{b}\, 
       \bar{q}[C_{7\gamma}R
\nonumber \\ 
&&\qquad\qquad
+C^\prime_{7\gamma}L]  
       \sigma_{\mu \nu }F^{\mu \nu }b+h.c.,  
\label{bqgamma}  
\end{eqnarray}  
where in the SM, $C_{7\gamma} \cong -0.3$ at   
the typical B decay energy scale $\mu \approx 5$ GeV.   
Due to the left-handed nature of weak interactions,   
$C_{7\gamma}$ dominates while $C^\prime_{7\gamma}$ 
is suppressed by $m_q/m_b$.
This may not be the case, however, in models beyond the SM,
and interesting $CP$ violating asymmetries in  
mixing induced radiative B decay can occur \cite{Atwood}. 
In a previous work \cite{b2sp},  
we have discussed the $b\to s \gamma$ process 
in the context of New Physics.   
We found that large direct $CP$ violations are possible, 
in contrast to SM expectations which are very small.  
For mixing induced $CP$ violation in B decay, 
we also find large asymmetries due to enhanced $C^\prime_{7\gamma}$.  
Furthermore, the mixing dependent asymmetry in $B\to\rho\gamma$  
is more accessible than in $B\to K^{*0}\gamma$,  
because $\rho^0\to\pi^+\pi^-$ can give the vertex information needed for time   
dependence, while $K^{*0}\to K_s\pi^0$ unfortunately does not \cite{Atwood}.  
Given that SM predictions on direct $CP$ violation in $b\to d\gamma$ are  
in general not small \cite{Ali},  
mixing induced $CP$ violation in $b\to d\gamma$   
is a more sensitive probe of New Physics.
Of course, mixing dependent $CP$ violation in $b\to s\gamma$ processes
can be more readily probed in $B_s$ system such as
in $B_s\to \phi\gamma$.

Mixing induced $CP$ violation in radiative B decay is similar 
to the golden $J/\psi K_S$ mode.  
The hadronic uncertainties factor out for self-conjugate $CP$ eigenstate(s).  
Furthermore, one needs both $B$ and $\bar B$ to decay to the same final state  
to allow interference to take place.   
Therefore, the formula of the asymmetry $a_{M^0\gamma}$,  
where $M^0$ is a $CP$ self-conjugate hadron   
such as $\rho^0,\,\omega,\,\phi,\,K^{*0}({\rm in}\,K_s\pi^0)$,  
resembles that of $a_{J/\psi K^0_s}$,  
\begin{equation}  
a_{M^0\gamma}=\xi \sin 2 {\vartheta}\,   
                     \sin[2\phi_1-\phi_{7\gamma}-\phi^{(\prime)}_{7\gamma}]\,  
                     \sin \Delta m\,t  
\label{amix}
\end{equation}  
where $\xi$ is the $CP$ eigenvalue of $M^0$,  
\begin{equation}  
\sin 2 {\vartheta}\equiv   
{{2\,|C_{7\gamma}C_{7\gamma}^{\prime }|}  
\over{|C_{7\gamma}|^{2}+|C_{7\gamma}^{\prime}|^{2}}},  
\label{s2theta}
\end{equation}  
is the mean strength of the two chiral amplitudes,
and $\phi^{(\prime)}_{7\gamma}$  
is the phase of $C^{(\prime)}_{7\gamma}$.  
While the $``\sin 2{\vartheta}"$ in the $B\to J/\psi K_S$ case is equal to one,
an important feature for $B\to M^0\gamma$ is that   
one needs both $C_{7\gamma}$ and $C^\prime_{7\gamma}$ for the interference  
to occur.  
This is the reason why these asymmetries are suppressed in the SM
since $|C^{\prime\rm SM}_{7\gamma}/C^{\rm SM}_{7\gamma}|\ll 1$,  
hence precisely why they are sensitive to New Physics.
The Belle Collaboration may be able to test the asymmetry to an accuracy
of 10\% in about a decade \cite{Hazumi}.
Hadron machines may also be able to study this
if they can observe the radiative B decay modes,
since they produce many more $B$$\bar B$s
than the $e^+\,e^-$ machines.

Another hint for New Physics may come from the rather large value of   
the newly observed $\varepsilon^\prime/\varepsilon$ \cite{KTEV}.
It has prompted New Physics considerations \cite{Masiero}, 
even though the experimental value could be accommodated within the SM. 
We do not consider New Physics hints from muon anomalous magnetic moment.

We are interested in New Physics models that can lead to 
deviations in the above mentioned processes.
It is interesting that supersymmetry (SUSY) models with 
Abelian horizontal symmetry (AHS) can provide 
a unified framework for all such New Physics effects. 
This will be presented in Section II,
where we will explain the implications of AHS on quark mixing,
in particular on the origin of large right-handed down quark mixings.
The effect is carried over to the squark sector in SUSY models,
and squark mixings will impact on flavor changing neutral currents (FCNC).
In face of stringent constraints from kaon mixings,
quark-squark alignment (QSA) is invoked to produce texture zeros,
where we will give an explicit example of horizontal charge assignments.
The subsequent sections are devoted to 
various FCNCs induced by squark mixings.
In Section III we study the effect on $B_d$--$\bar B_d$
and $B_s$--$\bar B_s$ mixings, 
which sets the scale of superparticle masses.
In Section IV, we show that the chargino contribution on 
kaon mixings gives a similar superparticle mass scale.
A generic feature of QSA is the possibility of 
sizable $D^0$--$\bar D^0$ mixing.
It is interesting that in AHS models with SUSY,
with sparticle scale fixed by $B$--$\bar B$ mixing,
$x_D$ could be right in the ball-park of the CLEO range,
as we show in Section V.
Section VI is devoted to radiative $B$ decay in SUSY models,
and discussions and conclusions are given in Sections VII and VIII, 
respectively.

\section{Abelian Flavor Symmetry 
	 with SUSY}  

\subsection{Abelian Flavor Symmetry and Large $\tilde d_R$ Mixings}  

Fermion masses and mixings in the 
Cabibbo-Kobayashi-Maskawa matrix $V_{\rm CKM}$ 
exhibit an intriguing hierarchical pattern  
in powers of $\lambda \equiv \vert V_{us}\vert$:
\begin{eqnarray}
m_u/m_c\sim\lambda^3,\quad
  m_c/m_t\sim\lambda^4,\quad
    m_t\sqrt{G_F}\sim 1,
\nonumber \\
m_d/m_s\sim\lambda^2,\quad
  m_s/m_b\sim\lambda^2,\quad
    m_b/m_t\sim\lambda^3,
\\
|V_{cb}|\sim\lambda^2,\qquad
    |V_{ub}|\sim\lambda^3.\qquad\quad\quad
\nonumber
\end{eqnarray}  
The structure could be due to an underlying symmetry \cite{horizontal},   
the breaking of which gives an expansion in   
$\lambda \sim \langle S\rangle/M$, with  
$S$ a scalar field and $M$ a high scale. 
In these models, after integrating out some massive fields of mass $M$, 
one obtains non-renormalizable terms \cite{horizontal,Nir93},
\begin{equation}
L_{\rm mass}=\lambda^q_{ij} H_d \Bigl({S\over M}\Bigr)^{\alpha_{q\,ij}} 
            Q_i \bar q_{Rj}+{\rm h.c.},
\label{Lmass}
\end{equation}
where $q$ are summed over up and down type quarks,
$\lambda^{q}$ are ${\cal O}(1)$ numbers, and $i,j$ are generation indices.
$L_{\rm mass}$ is made a horizontal symmetry singlet by choosing appropiate
powers of $S$, i.e. $\alpha_{q\,ij}$.
For models with Abelian horizontal symmetry,
without loss of generality \cite{Nir93}, we can define the horizontal charges
of the scalar fields as
\begin{equation}
H(H_d)=H(H_u)=0,\,H(S)=-1.
\end{equation} 
The breaking of the horizontal symmetry as well as electroweak symmetry lead to
quark mass elements,
\begin{equation}
M^*_{q\,ij}=\lambda^q_{ij} \langle H_q \rangle
\biggl({\langle S \rangle \over M}\biggr)^{\alpha_{q\,ij}},\,\,
\alpha_{q\,ij}=H(Q_i)+H(\bar q_{Ri}). 
\end{equation}
It is now easy to see that
\begin{equation}
M_{ij} M_{ji} \sim M_{ii} M_{jj}, \,\,
({i,\,j\, \rm not\,\, summed}),
\label{MijMji}
\end{equation}
which follows as a consequence of the commuting nature of horizontal
charges.

By assuming small mixing angles, as one tries to understand the hierarchical pattern
in $\lambda$ by AHS, 
the quark mass ratios fix the order of magnitude of the diagonal elements
of quark mass matrices.
The upper right part of the mass matrix $M_q$ corresponds to  
$U_{qL}$ rotation, which is related to $V_{\rm CKM}=U_{uL} U^\dagger_{dL}$.
Small mixing angle and naturalness imply $U_{qL}\sim V_{\rm CKM}$.
By using Eq. (\ref{MijMji}), one can work out the lower left part
and hence the whole mass matrix \cite{Nir93,Nir9394}  
\begin{equation}  
{M_u\over m_t} \sim \left[  
\matrix{\lambda^7 &\lambda^5 &\lambda^3 \cr  
         \lambda^6 &\lambda^4 &\lambda^2 \cr   
         \lambda^4 &\lambda^2 &1}  
\right],\quad  
{M_d\over m_b}\sim 
\left[  
\matrix{\lambda^4 &\lambda^3 &\lambda^3 \cr  
         \lambda^3 &\lambda^2 &\lambda^2 \cr   
         \lambda &1 &1}  
\right].  
\label{Mq}  
\end{equation}  
Since $U_{qL}$ are restricted by $V_{\rm CKM}$,  
mixing angles in $U_{qR}$ are in general greater.
In particular, we find that 
$M_d^{32}/m_b$ and $M_d^{31}/m_b$ are   
the most prominent off-diagonal elements.

To summerize, we have,
\begin{eqnarray}  
U_{qL}  
&\sim& {\left(  
\matrix{1 &\lambda &\lambda^3 \cr  
         \lambda &1 &\lambda^2 \cr   
         \lambda^3 &\lambda^2 &1}  
\right)},\quad  
U_{dR}  
\sim {\left(  
\matrix{1 &\lambda &\lambda \cr  
         \lambda &1 &1 \cr   
         \lambda &1 &1}  
\right)},
\nonumber \\
&&\quad\quad
U_{uR}  
\sim {\left(  
\matrix{1 &\lambda &\lambda^4 \cr  
         \lambda &1 &\lambda^2 \cr   
         \lambda^4 &\lambda^2 &1}  
\right)}.       
\label{Uq}
\end{eqnarray}
It is clear that mixing angles in $U_{dR}$ are in general greater than
those in $U_{qL}$ and $U_{uR}$, especially when the $b$ flavor is involved.
\footnote{
These large mixings in second and third generation $d_R$'s may be 
related to large mixing in second and third generation neutrinos
\cite{superK}
when considering grand unified theories (GUT) \cite{Elwood}.}
Although large mixings in the right handed down quark sector
are {\it useless} or well hidden within SM,    
$B_d$ and $B_s$ mixings are naturally susceptible to   
New Physics involving further dynamics related to
the right-handed down flavor sector.  

As one of the leading candidates for New Physics, SUSY  
helps resolve many of the potential problems that emerge when one   
extends beyond the SM, for example the gauge hierarchy problem,   
unification of SU(3)$\times $SU(2)$\times $U(1) gauge couplings,   
and so on \cite{SUSY}.
It is interesting to note that large mixing in right-handed 
down quark sector will be transmitted to right-handed down squarks,
if the breakings of flavor symmetry and SUSY are not closely related.
The flavor symmetry gives better control on soft breaking parameters
resulting in a more predictive SUSY model.
We now elevate Eq. (\ref{Lmass}) to the superpotential as well as 
similar forms in trilinear terms, the so called A terms.
By flavor symmetry, we still have the same power of $S$ 
to ensure that the whole term remains a horizontal singlet, 
that is,
\begin{equation}  
\label{MLR}
(\widetilde M^2_q)^{ij}_{LR}\sim\widetilde m\,M_d^{ij},\quad   
(\widetilde M^2_q)_{RL}=(\widetilde M^2_q)_{LR}^\dagger,
\end{equation} 
which are {\it roughly} proportional to respective quark mass matrices,  
hence their effects are suppressed by $m_q/\widetilde m$ \cite{Nir9394}.
While the symmetry does not require the new $\lambda_{ij}$ in the A-term
to be the same as $\lambda^q_{ij}$ in Eq. (\ref{Lmass}),
$(\widetilde M^2_q)_{LR}$ cannot, in general, be diagonal in the quark mass
basis.
From Eq.~(\ref{Mq}), one easily gets  
${(\widetilde M^2_Q)_{LL}/ \widetilde m^2} \sim~V_{\rm CKM}$, while  
\begin{equation}  
(\widetilde M^{2}_d)_{RR}  
\sim \widetilde m^2 \left[  
\matrix{1 &\lambda &\lambda \cr  
         \lambda &1 &1 \cr   
         \lambda &1 &1}  
\right].  
\label{MRR}      
\end{equation} 

The squark mixings will have impact on FCNC   
because of extra dynamics involving $\tilde q$-$\tilde g$, 
$\tilde q$-$\tilde\chi^\pm$ and $\tilde q$-$\tilde\chi^0$ couplings.   
It is now clear that FCNC processes involving $\tilde b_R$ are sensitive probes
of this generic class of SUSY models.
The $RR$ sector could   
contribute significantly to $B_d$ and $B_s$ mixings, to be discussed 
in the next section,
via large mixings in $\tilde b_R$--$\tilde d_R$ and in
$\tilde b_R$--$\tilde s_R$ as shown in Eq. (\ref{MRR}).

\subsection{Quark-squark Alignment}

In order to compare squark mixing angles with FCNC constrains,
we will use the mass insertion approximation \cite{MI,Gabbiani}
in the following discussion.
It is customary to take squarks as   
almost degenerate at scale $\widetilde m$.  
In quark mass basis, one defines \cite{MI},
\begin{equation}
\delta^{ij}_{qAB} \equiv  
[U^\dagger_{qA}\,(\widetilde M^2_q)_{AB}\, U_{qB}]^{ij}  
                         /{\widetilde m}^2,  
\label{delta}  
\end{equation}  
which is roughly the squark mixing angle,
$\widetilde M^2_q$ are squark mass matrices,  
$A,B=L,R$, and $i,j$ are generation indices. 
Note that $\delta^{13}_{dRR}\sim\lambda$ and 
$\delta^{23}_{dRR} \sim 1$,  
while LR and RL mixings are suppressed by $m_q/\widetilde m$.

It is well known that kaon mixings give stringent constraint on 
new flavor violating source.
The $12,21$ elements in $\widetilde M^2_Q$ and $(\widetilde M^2_d)_{RR}$ 
in Eq. (\ref{MRR}) will induce too large a contribution to kaon mixing
via gluino box diagrams \cite{Silvestrini,kaon}.
One has to suppress these squark mixings by  
enforcing approximate ``texture zeros".  
This can be done by invoking quark-squark alignment (QSA)
\cite{Nir93,Nir9394}, by using two (or more) singlet fields $S_i$ to   
break the $U(1)\times U(1)$ (or higher) Abelian horizontal symmetry,  
and making use of the holomorphic nature of   
the superpotential in SUSY models.
For example, we may have a term with negative power $\alpha_{qij}$ in
$S$ in Eq. (\ref{Lmass}) to satisfy the horizontal symmetry.
The term $S^{-|\alpha_{qij}|}$ is simply $(S^*)^{|\alpha_{qij}|}$.
As we promote Eq. (\ref{Lmass}) to superpotential,
which can only be a function of superfields and not of 
conjugate superfields at the same time, 
one can no longer use $S^*$ and hence there is
a zero in that particular $ij$-th element. 
One can have $M_d^{12,21}=0$ which imply $U^{12}_{dL,R}=0$ or 
are highly suppressed,  
and likewise $(\widetilde M^2_d)_{LL,RR}^{12}$ are also suppressed.  
Thus, $\delta^{12}_{dLL,RR}$ can be suppressed  
and the kaon mixing constraint is satisfied accordingly.   

There is one subtlety involving our  
choice to retain $(\widetilde M^2_d)_{RR}^{13}$,  
which arises from $M_d^{31}$.  
The mass matrix $M_d$ is diagonalized by a bi-unitary transform,    
hence, it is of the form
\begin{eqnarray}
{M_d\over m_b}&=& U^\dagger_{dL}\, {M^{\rm diag}_d\over m_b}\, U_{dR}
\nonumber \\
&\sim&{\left(  
\matrix{1                 &\lambda^a  &\lambda^b\cr  
        \lambda^a   &1                &\lambda^c \cr   
        \lambda^b     &\lambda^c  &1}  
\right)}\,
{\left(  
\matrix{\lambda^4  &0                &0 \cr  
        0                &\lambda^2  &0 \cr   
        0                &0                &1}
\right)}\,
{\left(  
\matrix{1                 &\lambda^d  &\lambda^e\cr  
        \lambda^d     &1                &\lambda^f \cr   
        \lambda^e     &\lambda^f  &1}  
\right)}
\label{biunitary}
\nonumber \\
\end{eqnarray}
where the diagonal matrix in the middle of the right hand side
corresponds to 
the diagonal down quark mass (ratio) matrix, 
and the one to its left (right) corresponds to $U_{dL}$ ($U_{dR}$). 
Multiplying out the matrices in Eq. (\ref{biunitary}), we have 
\begin{eqnarray}
{M_d^{11}\over m_b}\hspace{-0.2cm}&\sim&
                \lambda^4+\lambda^{2+a+d}+\lambda^{b+e},
\,\,
{M_d^{12}\over m_b}\hspace{-0.2cm}\sim
                \lambda^{4+d}+\lambda^{2+a}+\lambda^{b+f},
\nonumber \\
{M_d^{21}\over m_b}\hspace{-0.2cm}&\sim&
                \lambda^{4+a}+\lambda^{2+d}+\lambda^{c+e},
\,\,
{M_d^{22}\over m_b}\hspace{-0.2cm}\sim
                \lambda^{4+a+d}+\lambda^{2}+\lambda^{c+f},
\nonumber \\
{M_d^{31}\over m_b}\hspace{-0.2cm}&\sim&
                \lambda^{4+b}+\lambda^{2+c+d}+\lambda^{e},
\,\,
{M_d^{32}\over m_b}\hspace{-0.2cm}\sim
                \lambda^{4+b+d}+\lambda^{2+c}+\lambda^{f},
\nonumber \\
{M_d^{13}\over m_b}\hspace{-0.2cm}&\sim&
                \lambda^{4+e}+\lambda^{2+a+f}+\lambda^{b},
\,\,
{M_d^{23}\over m_b}\hspace{-0.2cm}\sim
                \lambda^{4+a+e}+\lambda^{2+f}+\lambda^{c},
\nonumber \\
{M_d^{33}\over m_b}\hspace{-0.2cm}&\sim&
                1+\lambda^{4+b+e}+\lambda^{2+c+f}.
\end{eqnarray}
We see that, by retaining $M_d^{31}/m_b \sim \lambda$,
we have $e=1$ and hence $U^{13}_{dR} \sim \lambda$.
We will have $c=2$, if we also retain $M_d^{23}/m_b\sim\lambda^2$.
However, as the kaon mixing constraint requires 
\begin{equation}
M_d^{21}/m_b\sim
\lambda^{4+a}+\lambda^{2+d}+\lambda^{c+e}\sim0,
\end{equation}
which in turn gives $d=1$ for $\lambda^{2+d}$ to be of the same order 
as $\lambda^{c+e}=\lambda^3$, to allow cancellation to take place.
This will still give squark mixing angle $\sim\lambda$ 
between $\tilde d_R$--$\tilde s_R$, which is not acceptable.
A closer look reveals that $M_d^{32}/m_b\sim 1$ is also unacceptable.
It gives $f=0$ from $M^{32}/m_b\sim 1$. 
With the requirement of $M_d^{23}=0$, we again have $c=2$
and it follows that $d=1$ by the same argument. 
Because $(\widetilde M^2_d)_{RR}^{13}/\widetilde m^2  
 \sim \lambda$ is kept,  
we need to make $M_d^{23}/m_b$ {\it and} $M^{32}_d/m_b$ also vanishing in 
face of stringent $\Delta m_K$ and $\varepsilon_K$ constraints.
The decoupling of $s$ flavor from other generations 
thus follows from imposing QSA in 12 sector {\it and}
choosing to retain $M^{31}_d\neq 0$.

In the usual approach of quark-squark alignment,  
this subtlety does not arise because  
one aspires to lower $m_{\tilde q}$, $m_{\tilde g}$  
for sake of collider and other signatures.   
Since $\delta^{13}_{dRR}\sim\lambda$ and $\delta^{23}_{dRR} \sim 1$  
would then violate $B_d$ mixing and $b\to s\gamma$ constraints already,  
they are eliminated from the outset.  
As a result, $M_d^{23}/m_b \sim \lambda^2$,  
though of little consequence, can be retained.

\subsection{Explicit Examples of QSA}

To be specific, we now give an explicit assignment of the horizontal charges of   
quark superfields and the resulting mass matrices as an illustrative example.  

Since $|V_{ub}|\sim 0.002-0.005<\lambda^3$ \cite{HouWong}, we may use 
a smaller parameter $\tilde \lambda=0.18$ instead of $\lambda$. 
We use two $S_i$ fields to break the horizontal symmetry,
\begin{equation}   
{\langle S_1\rangle\over M}\sim{\tilde\lambda}^{0.5}, \quad 
{\langle S_2\rangle\over M}\sim{\tilde\lambda}^{0.5}. 
\end{equation}  
The horizontal charges of $S_1$ and $S_2$ are $(-1,0)$, $(0,-1)$,   
respectively   
and those of $Q$, $\bar d_R$ and $\bar u_R$ are given by  
\begin{eqnarray}  
\label{horizontal}  
Q_1&:&\,(8,-2),\quad Q_2:\,(1,3),\quad Q_3:\,(2,-2),  
\nonumber \\  
\bar d_{R1}&:&\,(-2,10),\quad \bar d_{R2}:\,(9,-3),\quad \bar d_{R3}:\,(-2,8),  
\\  
\bar u_{R1}&:&\,(-3,11),\quad \bar u_{R2}:\,(0,3),\quad \bar u_{R3}:\,(-1,2),  
\nonumber  
\end{eqnarray}  
for small $\tan\beta$. For $\tan\beta\sim50$ we change horizontal charges  
of $\bar d_{Ri}$ to,  
\begin{equation}  
\bar d_{R1}:\,(-2,5),\quad \bar d_{R2}:\,(4,-3),\quad \bar d_{R3}:\,(-2,3).  
\end{equation}  
In this way we get,  
\begin{equation}  
{M_u\over m_t}\sim \left(  
\matrix{\tilde\lambda^{6.5} &\tilde\lambda^4 &\tilde\lambda^3\cr  
	 0                  &\tilde\lambda^3 &\tilde\lambda^2 \cr  
	 0                  &\tilde\lambda   &1}  
		  \right)  
,\quad  
{M_d\over m_t\,\tilde\lambda^{2.5}}\sim  \left(  
\matrix{\tilde\lambda^4  &0                &\tilde\lambda^3\cr  
         0               &\tilde\lambda^2  &0 \cr   
         \tilde\lambda   &0                &1}  
\right)  
.  
\label{mqnew}  
\end{equation}

From Eq. (\ref{mqnew}), 
we have effectively decoupled the second generation from the
first and third in $M_d$,
which corresponds to suppressed $U^{12,23}_{dL,dR}=0$.
The case is reminiscent of~\cite{b2sp}  
where we decouple $d$ flavor. 
The corresponding squark mass matrices from
Eq. (\ref{horizontal}) are   
$(\widetilde M^2_q)^{ij}_{LR}\sim\widetilde m\,M^{ij}_q$  
and   
\begin{eqnarray}  
(\widetilde M^2_Q)^{ij}_{LL}  
&\sim& \widetilde m^2\left(  
\matrix{1        &\tilde\lambda^6  &\tilde\lambda^3 \cr  
       \tilde\lambda^6 &1          &\tilde\lambda^3 \cr   
       \tilde\lambda^3 &\tilde\lambda^3  &1}  
\right),      
\\  
(\widetilde M^2_u)^{ij}_{RR}  
&\sim& \widetilde m^2\left(  
\matrix{1        &\tilde\lambda^{4.5}  &\tilde\lambda^{4.5} \cr  
       \tilde\lambda^{4.5} &1          &\tilde\lambda \cr  
	      \tilde\lambda^{4.5} &\tilde\lambda  &1}  
	      \right),  
\\ 
(\widetilde M^2_d)^{ij}_{RR}  
&\sim& \widetilde m^2\left(  
\matrix{1        &\tilde\lambda^{12}  &\tilde\lambda \cr  
       \tilde\lambda^{12} &1          &\tilde\lambda^{11} \cr  
      \tilde\lambda &\tilde\lambda^{11}  &1}  
      \right).     
\end{eqnarray}  
For large $\tan\beta$, we change   
$(\widetilde M^2_d)^{12}_{RR}$ and $(\widetilde M^2_d)^{23}_{RR}$ to  
$\tilde\lambda^7$ and $\tilde\lambda^6$, respectively.  
We summarize all $\delta$'s of interest in Table I.
We will see that these values are all well below   
the limits from $\Delta m_K$ and  
$\varepsilon$ constraints, even with $\cal O$(1) phases.
   
At this point one thing needs to be emphasized.
In face of severe kaon mixing constraints,
we did not choose horizontal charges to {\it create} large right 
handed $\tilde d$--$\tilde b$ squark mixings.
Instead, the choice of horizontal charges were to {\it retain} 
this natural large mixing that follow from the Abelian nature of the
underlying flavor symmetry. 
Both the Abelian flavor symmetry and the mixing pattern originate from
the observed mass mixing hierarchy pattern.
Phenomenological consequences of these large mixings \cite{Dudas}
should be explored \cite{ChuaHou}.

\begin{table}[b!]  
\label{AHS}
\twocolumn[\hsize\textwidth\columnwidth\hsize\csname
@twocolumnfalse\endcsname  
\begin{center}  
\begin{tabular}{c llll}  
$(q,ij)$  
&$|\delta^{ij}_{qLL}|$ 
&$|\delta^{ij}_{qRR}|$ 
&$|\delta^{ij}_{qLR}|$ 
&$|\delta^{ij}_{qRL}|$ 
\\ \hline 
(d,12) 
&$\tilde\lambda^6$    
&$\tilde\lambda^{12[7]}$  
&$m_b/\msq\,\tilde\lambda^8(\underline1+\{\tan\beta\})$  
&$m_b/{\widetilde m}\,(\underline{\tilde\lambda^{14[7]}}+\{\tilde\lambda^{4} 
\tan\beta\})$  
\\ 
(d,13)&$\tilde\lambda^3$&$\tilde\lambda$  
&$m_b/{\widetilde m}\, \tilde\lambda^{3}(1+\{\tan\beta\})$  
&$m_b/{\widetilde m}\, \tilde\lambda\,(1+\{\tan\beta\})$  
\\ 
(d,23)&$\tilde\lambda^3$&$\tilde\lambda^{11[6]}$  
&$m_b/{\widetilde m}\, \tilde\lambda^{3} ({\underline 1}+\{\tan\beta\})$  
&$m_b/{\widetilde m}\, (\underline {\tilde\lambda^{11[7]}}
+\{\tilde\lambda^{5} \tan\beta\})$  
\\ 
(u,12)&$\tilde\lambda$&$\tilde\lambda^{4.5}$  
&$m_t/{\widetilde m}\, \tilde\lambda^{4}$  
&$m_t/{\widetilde m}\, (\underline{\tilde\lambda^{8.5}}+\{\tilde\lambda^{7.5}\})$  
\\ 
(u,23)&$\tilde\lambda^2$&$\tilde\lambda$  
&$m_t/{\widetilde m}\, \tilde\lambda^{2}$  
&$m_t/{\widetilde m}\, \tilde\lambda$  
\end{tabular}  
{\caption{The order of magnitudes of $|\delta_q^{ij}|$s  
from the Abelian horizontal symmetry model, where  
$\tilde\lambda=0.18$. 
Terms in parentheses [...] correspond to $\tan\beta\sim50$ case  
and terms with \{...\} only exist when we consider non-standard soft breaking  
terms.
The underlined terms are from filled zeros (see text).  
}}  
\end{center}
]  
\end{table} 

There is a generic feature \cite{Nir93,Nir9394} of QSA 
that is worthy of note. Having $U^{12}_{dL} \simeq 0$ implies 
$U^{12}_{uL}\sim \vert V_{cd}\vert =\lambda$, 
which can also be read off from Eq. (\ref{mqnew}).
One now has $\delta^{12}_{uLL} \sim \tilde\lambda$,
as one can see from Eq. (\ref{delta}).
This Cabibbo strength $\delta^{12}_{uLL}$
can contribute to kaon mixing via chargino diagrams, 
and also $D^0$--$\bar D^0$ mixing via gluino diagrams,
as we will dicuss in Sec. IV and V, respectively. 
We note that New Physics contributions to 
$D^0$--$\bar D^0$ mixing are of great interest at present, 
since the recent CLEO (and FOCUS) results
may be a hint for $D^0$ mixing in disguise.
Note also that the texture zeros of $M_u^{21,31}$ are 
generated through QSA.
The zero of $M_u^{21}$ is needed to avoid
$\delta^{12}_{uRR}\sim\tilde\lambda$, for otherwise,
together with $\delta^{12}_{uLL}\sim\tilde\lambda$ they will induce
too large a contribution to $D$ mixing. 
The zero of $M_u^{31}$ is to avoid a feed back to $\delta^{12}_{uRR}$, 
analogous to the discussion in the previous subsection.

We mention another subtlety arising from the K\"ahler potential \cite{Dudas}.
When the horizontal symmetry is spontaneously broken,
mixing also occurs in the kinetic terms.
By canonical normalization of the kinetic terms, further mixings are introduced.
For example, $M_q$ now becomes
$L_q\, M_q \, R^\dagger_q$, where
$L_q\sim(\widetilde M^2_Q/\widetilde m^2)^{-1/2}$ and
$R^\dagger_q\sim(\widetilde M^2_{qRR}/\widetilde m^2)^{-1/2}$.
The zeros in Eq. (\ref{mqnew}) are now all lifted, 
and are called filled zeros \cite{Dudas},
giving rise to the underlined terms in Table I.
$U_{qL}$ also becomes $L_q\, U_{qL}$
and similarly for other mixing matrices.
Modifications of previous results can be achieved by suitable rotations
and are also shown in Table I.  

A second possibility of horizontal charge assignment
is to retain $M_d^{32}$ and $M_d^{23}$ 
while having vanishing $M_d^{31}$,
i.e.
\begin{equation}
\label{Mdb2sp}
{M_d\over m_b}\sim 
\left( \matrix{\lambda^4       &0           &0\cr  
               0               &\lambda^2   &\lambda^2 \cr   
               0               &1           &1}  
\right).  
\end{equation}
The assignment of horizontal charges for this case
can be found in Ref. \cite{Nir93}.
The squark mixing can generate sizable contribution in $B_s$ mixings
\cite{Barenboim}.
In this case one may have large CP phase in $B_s$ mixing and 
possible effects in $b\rightarrow s\gamma$ to be discussed later.
The democratic structure of $\widetilde M^2_{dRR}$ in Eq. (\ref{MRR})
in the 2-3 sub-matrix leads to 
approximate maximal mixing in $\tilde s_R$ and $\tilde b_R$.
The large off-diagonal elements $(\widetilde M^2_d)^{23,32}_{RR}$
lead to large level splitting.
This allows for a possibly light strange beauty squark 
with interesting impact in $B_s$ mixing and direct search,
but leaving Br$(B\to X_s\gamma)$ largely unaffected \cite{sb}.  

It is interesting that, faced with stringent kaon constraint, 
AHS models with QSA allow
large mixing in either $\tilde b_R-\tilde d_R$ 
or $\tilde b_R-\tilde s_R$, but not both at the same time.
Thus, a prediction of this model is,
if it is responsible for the smallness of the measured $\sin 2\phi_1$ 
(assuming that the low value persists in the future),
there will be no large New Physics contribution to $B_s$ mixing.

We now study the FCNC induced by these squark mixings
in the following sections.

\section{$B^0$--$\bar B^0$ Mixing}  
                                 
In this section, we first focus on the general formalism of
neutral $B$ meson mixings in the AHS model with SUSY.
We will focus on the $B_d$ system for applications,
which is readily extendable to the $B_s$ system.
For the latter system, the intriguing possibility
that large right-handed squark mixings could lead to 
a light ``strange-beauty" squark will be discussed briefly
in Sec. VII.C.

\begin{figure}[t!]
\centerline{\epsfxsize 8 cm \epsffile{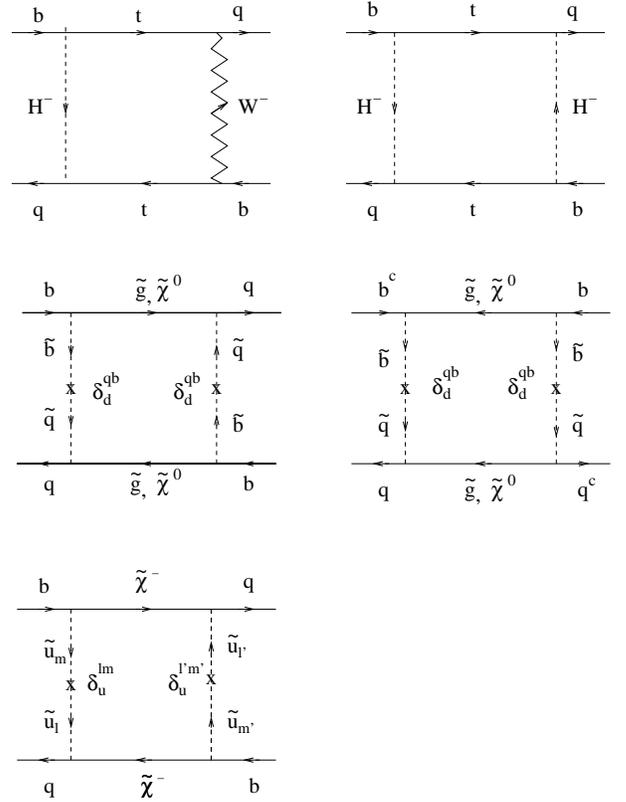}}
\smallskip
\caption{SUSY box diagrams for $\Delta B=2$ processes.   
}
\label{susybox}
\end{figure}

The effective Hamiltonian for $B^0_q$--$\bar B^0_q$ mixings  
from SUSY contributions, where $q=d$ or $s$, 
is given by   
\begin{equation}  
H_{\rm eff}={\sum_i C_i {\cal O}_i},   
\label{Heff}  
\end{equation}  
where,  
\begin{eqnarray}  
&{\cal O}_1=\bar q^\alpha_{L}\gamma_\mu b^\alpha_L\,  
           \bar q^\beta_{L}\gamma^\mu b^\beta_L,  
\nonumber \\  
{\cal O}_2&=\bar q^\alpha_{L} b^\alpha_R\,  
           \bar q^\beta_{L} b^\beta_R,  
\quad  
{\cal O}_3=\bar q^\alpha_{L} b^\beta_R\,  
             \bar q^\beta_{L} b^\alpha_R,  
\\  
{\cal O}_4&=\bar q^\alpha_{L} b^\alpha_R\,  
           \bar q^\beta_{R} b^\beta_L,  
\quad  
{\cal O}_5=\bar q^\alpha_{L} b^\beta_R\,  
             \bar q^\beta_{R} b^\alpha_L,  
\nonumber  
\end{eqnarray}  
together with three other operators ${\tilde {\cal O}}_{1,2,3}$   
that are chiral conjugations ($L\leftrightarrow R$) of ${\cal O}_{1,2,3}$.  
The Wilson coefficients receive charged Higgs, chargino, gluino,
gluino-neutralino, and neutralino exchange box diagram contributions,
\begin{equation}  
C_i=C^{H^-}_i+C^{\tilde \chi^-}_i+C^{\tilde g}_i+  
    C^{\tilde g \tilde \chi^0}_i+C^{\tilde \chi^0}_i,
\end{equation}  
where the Feynman diagrams are shown in Fig. \ref{susybox}.

\subsection{Formulas}

\noindent \underline {Charged Higgs box}\cite{Bertolini,Branco}:  
\begin{eqnarray}  
C^{H^-}_1&=&{\alpha_{\rm w}^2\over 8 m_W^2} (V_{tb} V^*_{tq})^2   
          \bigg[x_{t\rm W}\, x_{tH} \cot^4 \beta {1\over 4} G(x_{tH},x_{tH})  
\nonumber \\             
&&\qquad+2 x_{t\rm W}^2 \cot^2 \beta   
 \bigg(F^\prime(x_{t\rm W},x_{t\rm W},x_{H\rm W})
\nonumber \\             
&&\qquad+{1\over 4}G^\prime(x_{t\rm W},x_{t\rm W},x_{H\rm W})\bigg)\bigg],  
\label{CH} \\  
C^{H^-}_2&=&-{\alpha_{\rm w}^2\over 8 m_W^2} (V_{tb} V^*_{tq})^2  
           x_{b\rm W}\big[x_{t\rm W} x_{tH} F(x_{tH},x_{tH}) 
\nonumber \\             
&&\qquad+2 x_{t\rm W}\,x_{t\rm W}\,\cot\beta\,
F^\prime (x_{t\rm W},x_{t\rm W},x_{H\rm W})\big],  
\nonumber  
\end{eqnarray}  
where $x_{ab}\equiv m_a^2/m_b^2$, and the loop functions 
$F^{(\prime)},G^{(\prime)}$ are given in Ref. \cite{Bertolini}.  
The charge Higgs contributions to   
other Wilson coefficients are either vanishing or 
suppressed by $m_q/m_W$.  

\vspace{0.5cm}
\noindent \underline{Chargino box}:
\begin{eqnarray}  
C^{\tilde \chi^-}_1 &=& {\alpha_{\rm w}^2\over \,\,\,8 m^2_{\tilde\chi^-_k}}  
   \,{\widetilde G^\prime(x_{\tilde q \tilde\chi^-_k},  
                        x_{\tilde \chi^-_j \tilde\chi^-_k})}  
{\cal A}_{jk}\,{\cal A}_{kj}
\nonumber \\  
C^{\tilde \chi^-}_3 &=&-{\alpha_{\rm w}^2\over \,\,\,2 m^2_{\tilde\chi^-_k}}  
\sqrt{x_{\tilde \chi^-_j \tilde \chi^-_k}}  
{\widetilde F^\prime(x_{\tilde q \tilde\chi^-_k},  
                        x_{\tilde \chi^-_j \tilde\chi^-_k})}
\nonumber \\
&&\qquad\qquad\qquad{\cal U}_{j2}\,{\cal U}_{k2}\,\hat Y^2_b {\cal B}_j{\cal B}_k
\label{CX}   
\end{eqnarray}
where the indices $j,k$ are summed over 1 to 2, and
\begin{eqnarray}
{\cal A}_{jk}&\equiv&{\cal V}_{j1} {\cal V}_{k1}^* V^*_{lq} V_{mb} \delta^{lm}_{uLL}
-{\cal V}_{j2} {\cal V}_{k1}^* \hat Y_t V^*_{tq} V_{tb} \delta^{33}_{uRL}
\nonumber \\
&&
-{\cal V}_{j1} {\cal V}_{k2}^* \hat Y_t V^*_{tq} V_{tb} \delta^{33}_{uLR}
+{\cal V}_{j2} {\cal V}_{k2}^* \hat Y_t^2 V^*_{lq} V_{mb} \delta^{33}_{uRR},
\nonumber \\ 
{\cal B}_j&\equiv&{\cal V}_{j1} V^*_{lq} V_{mb} \delta^{lm}_{uLL}
  -{\cal V}_{j2}\hat Y_t V^*_{tq} V_{tb} \delta^{33}_{uRL},
\end{eqnarray} 
the indices $l,m$ are   
summed over 3 generations of up type squarks,
$\hat Y_{u,c,t}=m_{u,c,t}/(\sqrt 2 m_W \sin\beta)$ and similarly,   
      $\hat Y_{d,s,b}=m_{d,s,b}/(\sqrt 2 m_W \cos\beta)$.  
The chargino mixing matrices ${\cal U},{\cal V}$ in Eq. (\ref{CX}) 
diagonalize the chargino
mass matrix,
\begin{equation}
M_{\widetilde\chi^{\pm}}={\cal U}^* 
\left(
\begin{array}{cc}
M_2 &\sqrt 2 m_W \sin\beta \\
\sqrt 2 m_W \cos\beta &\mu\\
\end{array}
\right)
{\cal V}^\dagger,
\end{equation}
and  
\begin{eqnarray}  
&&\Bigl(\widetilde F^\prime(x,y),\,\widetilde G^\prime(x,y)\Bigr)
\nonumber \\  
&&=x^2 \partial_a \partial_b \Bigl(F^\prime(a,b,y),  
                                G^\prime(a,b,y)\Bigr) \Bigr|_{a=b=x},
\end{eqnarray}
which can also be expressed as
\begin{equation}
\left(\widetilde F^\prime(x,y),\,\widetilde G^\prime(x,y)\right)
=\int_0^\infty  {dk^2\,x^2 (k^2)^{(1,2)}\over (k^2+1) (k^2+x)^4 (k^2+y)},
\label{FG}
\end{equation} 
which are always positive.
There is no chargino contribution to $C_2$ and $\tilde C_2$
because of the color structure of the chargino box diagrams,  
and other terms are suppressed by the smallness of $\hat Y_q$.
Since $\hat Y_t$ is large and $\hat Y_b$ can also be large for 
the case of large $\tan\beta$,
we keep them in $C^{\tilde \chi^-}_{1,3}$.
Note that in the AHS models, 
$\sum_{l,m} V^*_{lq} V_{mb} \delta^{lm}_{uLL}$ and 
$V^*_{tq} V_{tb} \delta^{33}_{uRR}$ are roughly of the order $|V_{tq}|$.
As we will show soon, the $\Delta m_{B_d}$ constraint require $\msq\sim$ TeV 
due to large $\tilde b_R$--$\tilde d_R$ mixings.
A typical LR stop mixing term contains $\hat Y_t V^*_{tq} \delta^{33}_{uLR}
\sim 1.7\, V^*_{tq} m_t/\tilde m$, which will be as small as $\sim 0.1 V^*_{tq}$
for $\tilde m\sim$ TeV.
Furthermore, the flavor scale may not be too far from TeV \cite{Nir93},
and there may not be much room for RG running to bring down the stop mass.
Unlike the usual approach where one has light stop,
the contributions from stop LR mixings are relatively small here.

\vspace{0.5cm}
\noindent \underline{Gluino box} \cite{Gabbiani}:  
\begin{eqnarray}
C^{\tilde g}_1&=&{\alpha^2_s\over \msq^2}\biggl[
{1\over 4}\biggl(1-{1\over N_c}\biggr)^2\,\xgq f_6(\xgq) 
\nonumber \\        
&&\qquad +{1\over 8}\biggl(N_c-{2\over N_c}+{1\over N_c^2}\biggr)
          \tilde f_6(\xgq)\biggr]\,(\delta^{q3}_{dLL})^2,  
\nonumber \\
C^{\tilde g}_4&=&{\alpha^2_s\over \msq^2}
\biggl[\biggl(N_c-{2\over N_c}\biggr)\,\xgq f_6(\xgq)  
                      -{\tilde f_6(\xgq)\over N_c} \biggr]\,  
                \delta^{q3}_{dLL}\delta^{q3}_{dRR},  
\nonumber \\  
C^{\tilde g}_5&=&{\alpha^2_s\over \msq^2}
\biggl[{\xgq f_6(\xgq)\over N_c^2}  
                      +\biggl({1\over 2}+{1\over 2 N_c^2}\biggr)
                                           \tilde f_6(\xgq)\biggr]\,  
                \delta^{q3}_{dLL}\delta^{q3}_{dRR},  
\nonumber\\    
\label{Cg}  
\end{eqnarray}  
where $N_c$ is the number of colors, and
\begin{eqnarray}  
f_6(x)&=&{(17-9x-9x^2+x^3+6\ln x+18x\ln x)\over 6(x-1)^5} ,  
\nonumber \\  
\tilde f_6(x)&=&{(1+9x-9x^2-x^3+6x\ln x+6x^2\ln x)\over 3 (x-1)^5} .  
\end{eqnarray}
They are related to
$\widetilde F^\prime(x,y)$ and $\widetilde G^\prime(x,y)$ by
\begin{equation}
\left(x f_6(x),\,-\tilde f_6(x) \right)=
x^{-1} \left({\widetilde F^\prime(x^{-1},1)},
\,{\widetilde G^\prime(x^{-1},1)}\right).
\end{equation}  
$\tilde C^{\tilde g}_1$ is obtained by interchanging 
$L\leftrightarrow R$ in $C^{\tilde g}_1$.
We neglect $C^{\tilde g}_{2,3}$ and $\tilde C^{\tilde g}_{2,3}$ 
due to the smallness of LR and RL mixings.
There are usual box and crossed box diagrams.
Terms with $N_c$ are from the former 
while ${\cal O}(1)$ terms are from the latter, 
which can be easily checked by 't~Hooft double line notation.
Terms with $1/N_c,\,1/N_c^2$ are subleading contributions 
from these two types of diagrams. 
We note that $C^{\tilde g}_4$ contains the largest $N_c$ factor 
hence is the most sensitive to squark mixings.
Note also that one has a zero in 
$C^{\tilde g}_1$ ($\tilde C^{\tilde g}_1$) for $\xgq\sim 2.43$.

\vspace{0.5cm} 
\noindent \underline{Gluino-neutralino box}:  
\begin{eqnarray}  
C^{\tilde g\tilde \chi^0}_1&=&  
{\alpha_s\alpha_{\rm w}\over 2 \tilde m^2_{\tilde g} }
   \biggl(1-{1\over N_c}\biggr)   
   \biggl[G_L^j G_L^{*j}   
   \tilde G^\prime(x_{\tilde q \tilde g},x_{\tilde \chi^0_j \tilde g})
\nonumber \\ 
&&
\hspace{-0.4cm}
-(G^j_L G^j_L+G^{*j}_L G^{*j}_L)
 \sqrt {x_{\tilde \chi^0_j \tilde g}}  
   \tilde F^\prime(x_{\tilde q \tilde g},x_{\tilde \chi^0_j \tilde g})   
          \biggr]\,(\delta^{q3}_{dLL})^2,  
\nonumber \\  
C^{\tilde g\tilde \chi^0}_2&=&  
{\alpha_s\alpha_{\rm w}\over 2 m^2_{\tilde g}}
   \biggl(1-{1\over N_c}\biggr)   
          H^j_{bL} H^j_{bL}
\nonumber\\
&&\qquad \times\sqrt {x_{\tilde \chi^0_j,\tilde g}}  
   \tilde F^\prime(x_{\tilde q \tilde g}, x_{\tilde \chi^0_j \tilde g})   
          \,(\delta^{q3}_{dLL})^2,  
\nonumber \\  
C^{\tilde g\tilde \chi^0}_3&=&
{\alpha_s\alpha_{\rm w}\over 2 m^2_{\tilde g}}
   \biggl(1-{1\over N_c}\biggr)   
          H^j_{bL} H^j_{bL}
\nonumber\\
&&\qquad\times\sqrt {x_{\tilde \chi^0_j,\tilde g}}  
   \tilde F^\prime(x_{\tilde q \tilde g}, x_{\tilde \chi^0_j \tilde g})   
          \,(\delta^{q3}_{dLL})^2,  
\nonumber \\  
C^{\tilde g\tilde \chi^0}_4&=&  
{\alpha_s\alpha_{\rm w}\over m^2_{\tilde g} }   
   \biggl[\biggl(G_R^j G_L^j+G_R^{*j} G_L^{*j}
                 -{1\over N_c}H^j_{bL} H^{j}_{bR}\biggr)  
\nonumber \\           
&&\qquad\quad\times\tilde G^\prime(x_{\tilde q \tilde g},x_{\tilde\chi^0_j \tilde g})
-2\,(G^j_L G^{*j}_R+G^{*j}_L G^j_R)
\nonumber \\
&&\qquad\quad\times\sqrt {x_{\tilde \chi^0_j \tilde g}}  
   \tilde F^\prime(x_{\tilde q \tilde g},x_{\tilde \chi^0_j \tilde g}\biggr)   
          \biggr]\,\delta^{q3}_{dLL}\delta^{q3}_{dRR},  
\nonumber \\  
C^{\tilde g\tilde \chi^0}_5&=&  
{\alpha_s\alpha_{\rm w}\over 2\tilde m^2_{\tilde g} }   
   \biggl[\biggl(H^j_{bL} H^j_{bR}-{2\over N_c}G_R^j G_L^j-
                                   {2\over N_c}G_R^{*j} G_L^{*j}\biggr)
\nonumber \\
&&\qquad\quad\times\tilde G^\prime(x_{\tilde q \tilde g},x_{\tilde\chi^0_j \tilde g})
+{4\over N_c}(G^j_L G^{*j}_R+G^{*j}_L G^j_R)
\nonumber \\
&&\qquad\quad \times\sqrt {x_{\tilde \chi^0_j\tilde g}}  
   \tilde F^\prime(x_{\tilde q\tilde g},x_{\tilde \chi^0_j\tilde g})   
          \biggr]\,\delta^{q3}_{dLL}\delta^{q3}_{dRR},  
\label{CgX0}  
\end{eqnarray}  
where j is summed over 1 to 4.
The mixing matrices $G_{L,R}, H_{bL,R}$ are given by  
\begin{eqnarray}  
&G^j_L\equiv \tan\theta_W Y_{Q} {\cal N}^*_{j1}  
         +T_{3D} {\cal N}^*_{j2},  
\,\,\,   
G^j_R&\equiv \tan\theta_W Q_d {\cal N}^*_{j1}  
\nonumber \\  
&\,\, H^j_{d,s,bL}\equiv{\cal N}_{j3} \hat Y_{d,s,b},  
\qquad\qquad 
H^j_{d,s,bR}&\equiv {\cal N}^*_{j3} \hat Y_{d,s,b},
\end{eqnarray}  
where $Y_{Q}$ is the usual hypercharge,
and the neutralino mixing matrix ${\cal N}$ diagonalizes 
the mass matrix $M_{\widetilde\chi^0}
={\cal N}^* {\cal M} {\cal N}^\dagger$,
\begin{eqnarray}
{\cal M}\hspace{-0.1cm}=\hspace{-0.1cm}\left(\hspace{-0.1cm}
\begin{array}{cccc}
M_1        &0         &-m_Z s_W c_\beta  &m_Z s_W s_\beta \\
0          &M_2       &m_Z c_W c_\beta  &-m_Z c_W s_\beta \\
-m_Z s_W c_\beta  &m_Z c_W c_\beta  &0 &-\mu\\ 
m_Z s_W s_\beta  &-m_Z c_W s_\beta  &-\mu &0\\ 
\end{array}
\hspace{-0.1cm}\right).
\nonumber 
\end{eqnarray} 
$\tilde C^{\tilde g\tilde \chi^0}_{1,2,3}$ are obtained by 
chiral conjugation.
Due to the smallness of $m_{d,s}$, we neglect terms with $H^j_{d,sL}$.  
Note that in Eq. (\ref{CgX0}), terms with $G^j G^{*j}$ are from the 
usual box diagram, while terms with $G^j G^j,\,G^{*j} G^{*j},\,H^j H^j$
are from the crossed diagram.
If the mass parameters $M_1,\ M_2,\ |\mu|\gg m_{\rm Z}$, 
we will have simpler forms for ${\cal N}$, where 
${\cal N}_{j1,j2}\sim\delta_{j1,j2}$, 
and higgsinos become maximally mixed.
This will lead to a large cancellation of higgsino contributions
in $C^{\tilde g\tilde \chi^0}_{2,3},\,\tilde C^{\tilde g\tilde \chi^0}_{2,3}$.
The diagonalization of the neutralino mass matrix usually leads to a
negative mass eigenvalue, say $m_{\tilde \chi^0_i}$.
It is well know that one can deal with it by two equivalent ways.
One could either choose the phases in ${\cal N}$ such that $m_{\tilde
\chi^0_i}$ are real and positive, 
or one could absorb the negative sign into $P_L \widetilde \chi^0_i$,
and modify Feynman rules accordingly \cite{SUSY,Gunion}.
However, there is a subtlety when dealing with the crossed diagrams
in the latter approach.
An additional negative sign is required for crossed box amplitudes when
$\widetilde \chi^0_i$ is in the loop, since 
$(\widetilde \chi^0_i)^c=-\widetilde \chi^0_i$ for that particular $i$.

\vspace{0.5cm}
\noindent \underline {Neutralino box} :  
\begin{eqnarray}  
C^{\tilde \chi^0}_1\hspace{-0.2cm}&=&\hspace{-0.1cm}  
{\alpha^2_{\rm w}\over 2 m^2_{\tilde\chi^0_k} }   
   \biggl[G_L^j G_L^{*j} G_L^k G_L^{*k}   
     \tilde G^\prime  
       (x_{\tilde q \tilde\chi^0_k},x_{\tilde\chi^0_j \tilde\chi^0_k})
\nonumber \\
&&  
  -2\,G^j_L G^j_L G_L^{*k} G_L^{*k}   
     \sqrt{x_{\tilde \chi^0_j \tilde\chi^0_k}}  
        \tilde F^\prime  
          (x_{\tilde q \tilde\chi^0_k},x_{\tilde\chi^0_j \tilde\chi^0_k})   
            \biggr]\,(\delta^{q3}_{dLL})^2,  
\nonumber \\  
C^{\tilde \chi^0}_2\hspace{-0.2cm}&=&\hspace{-0.1cm}  
{\alpha^2_{\rm w}\over m^2_{\tilde\chi^0_k} }   
   H^j_{bL} H^j_{bL} G_L^{*k} G_L^{*k}
\nonumber \\   
&&\times\sqrt{x_{\tilde \chi^0_j\tilde\chi^0_k}}  
      \tilde F^\prime  
       (x_{\tilde q \tilde\chi^0_k},x_{\tilde\chi^0_j \tilde\chi^0_k})   
          \,(\delta^{q3}_{dLL})^2,  
\nonumber \\
C^{\tilde \chi^0}_3\hspace{-0.2cm}&=&\hspace{-0.1cm}  
{\alpha^2_{\rm w}\over m^2_{\tilde\chi^0_k} }  
\sqrt{x_{\tilde \chi^0_j \tilde\chi^0_k}}  
\tilde F^\prime  
(x_{\tilde q \tilde\chi^0_k},x_{\tilde\chi^0_j \tilde\chi^0_k})   
\bigl(H^j_{bL} H^j_{bL} G_L^{*k} G_L^{*k}
\nonumber \\
&&-H^j_{bL} G^{*j}_L H^k_{bL} G^{*k}_L\bigr)   
          \,(\delta^{q3}_{dLL})^2,  
\label{CX0}\\  
C^{\tilde \chi^0}_4\hspace{-0.2cm}&=&\hspace{-0.1cm} 
 {\alpha^2_{\rm w}\over m^2_{\tilde\chi^0_k} }  
   \tilde G^\prime  
    (x_{\tilde q \tilde\chi^0_k},x_{\tilde\chi^0_j \tilde\chi^0_k})   
      \bigl(H^j_{bR} G^{*j}_L H^k_{bL} G_R^{*k} 
\nonumber \\  
&&+\,H^j_{bL} H^j_{bR} G^{*k}_L G^{*k}_R\bigr)   
          \,\delta^{q3}_{dLL} \delta^{q3}_{dRR},  
\nonumber \\  
C^{\tilde \chi^0}_5\hspace{-0.2cm}&=&\hspace{-0.1cm}  
{2\alpha^2_{\rm w}\over m^2_{\tilde\chi^0_k} }   
   \biggl[G_R^j G_L^j G_R^{*k} G_L^{*k}   
\tilde G^\prime  
(x_{\tilde q \tilde\chi^0_k},x_{\tilde\chi^0_j \tilde\chi^0_k})  
\nonumber \\  
&&-2\,G^j_R G^{*j}_L G^k_L G^{*k}_R
\nonumber\\
&&\times   
\sqrt{x_{\tilde \chi^0_j,\tilde\chi^0_k}}  
\tilde F^\prime  
(x_{\tilde q \tilde\chi^0_k},x_{\tilde\chi^0_j \tilde\chi^0_k})   
          \biggr]\,\delta^{q3}_{dLL} \delta^{q3}_{dRR},  
\nonumber  
\end{eqnarray}  
where the indices $j,k$ are summed over 1 to 4.
We make use of the fact that 
$\tilde G^\prime(x_{\tilde q \tilde\chi^0_k},x_{\tilde\chi^0_j \tilde\chi^0_k})
/m^2_{\tilde\chi^0_k}$ and 
$\sqrt{x_{\tilde \chi^0_j,\tilde\chi^0_k}}  
\tilde F^\prime  
(x_{\tilde q \tilde\chi^0_k},x_{\tilde\chi^0_j \tilde\chi^0_k})
/m^2_{\tilde\chi^0_k}$ are symmetric under $j\leftrightarrow k$,
which can be verified by using Eq. (\ref{FG}).  
$\tilde C^{\tilde\chi^0}_i$ are obtained by chiral conjugation.  
One can recognize contributions from the usual box or crossed box diagrams
by similar rules stated earlier.

We note that the
$C_1$s obtained in these four type of SUSY contributions are consistent
with those in Ref. \cite{Bertolini} by leading order Taylor expansion with 
respect to squark mixing angles. 

After obtaining these Wilson coefficients at SUSY scale $M_{\rm SUSY}$,
we apply renormalization group running to obtain 
$B^0$ mass scale values.
The renormalization group running of these Wilson coefficients   
including leading order QCD corrections is given by \cite{Bagger},  
\begin{eqnarray}  
C_1(\mu)&=&\eta_1 C_1(M_{\rm SUSY}),  
\nonumber \\  
C_2(\mu)&=&\eta_{22} C_2(M_{\rm SUSY})+\eta_{23} C_3(M_{\rm SUSY}),  
\nonumber \\  
C_3(\mu)&=&\eta_{32} C_2(M_{\rm SUSY})+\eta_{33} C_3(M_{\rm SUSY}),  
\\  
C_4(\mu)&=&\eta_{4} C_4(M_{\rm SUSY})+{1\over 3} (\eta_4-\eta_5) C_5(M_{\rm SUSY}),  
\nonumber \\  
C_5(\mu)&=&\eta_{5} C_5(M_{\rm SUSY}),  
\nonumber  
\end{eqnarray}   
where  
\begin{eqnarray}  
\label{etas}  
\eta_1&=&\biggl({\alpha_s(M_{\rm SUSY})\over \alpha_s(m_t)}\biggr)^{6/21}  
       \biggl({\alpha_s(m_t)\over \alpha_s(m_b)}\biggr)^{6/23},  
\nonumber \\  
\eta_2&=& \eta_1^{-2.42},\,  
\eta_3=\eta_1^{2.75},\,\eta_4=\eta_1^{-4},\, \eta_5=\eta_1^{1/2},  
\nonumber \\  
\eta_{22}&=&0.983\eta_2+0.017 \eta_3,\,  
\eta_{23}=-0.258\eta_2+0.258\eta_3,  
\nonumber \\  
\eta_{32}&=&-0.064\eta_2+0.064 \eta_3,\,  
\eta_{33}=0.017\eta_2+0.983\eta_3.  
\end{eqnarray}
It is clear that $\eta_4>\eta_2>1>\eta_5>\eta_1>\eta_3$.
Therefore, $C_2$ and, especially, $C_4$ will 
be enhanced by the RG running \cite{Bagger}.

\subsection{Impact on $B_d$ Mixing}

To obtain $\Delta m_{B_d}$, we use
$\Delta m_{B_d}=2 |M^B_{12}|$, where
\begin{eqnarray} 
M^B_{12} &\equiv& \vert M^B_{12}\vert \, e^{2i\Phi_{B_d}}  
\nonumber \\
 &=& \vert M^{\rm SM}_{12}\vert \, e^{2i\phi_1}  
 + \vert M^{\rm SUSY}_{12}\vert \, e^{i\phi_{\rm SUSY}},  
\label{MB12}
\end{eqnarray}  
and
\begin{equation}
M^{\rm SM}_{12}=0.33\, 
              \Biggl({f_{B_d}\sqrt{\hat B_{B_d}}\over 230\,{\rm MeV}}\Biggr)^2
\Biggl({|V_{td}|\over 8.8\times10^{-3}}\Biggr)^2 e^{2i\phi_1}\,\,
             {\rm ps}^{-1}, 
\nonumber \\ 
\end{equation}  
where $M^{\rm SM}_{12}$ is the SM contribution, its value is well  
known \cite{BurasLecture}.
The vacuum insertion matrix elements of ${\cal O}_i$ are 
given in Ref. \cite{Gabbiani}.
These matrix elements are modified by bag-factors to 
include non-factorizable effects.
For simplicity, we assume the bag-factors for matrix elements of
${\cal O}_{2-5}$ are equal to $\hat B_{B_d}$, which is calculated for 
${\cal O}_1$.
In the subsequent numerical analysis, 
we take $f_{B_d}\hat B_{B_d}^{1/2}=(230\pm 40)$ MeV \cite{fBd}.
For CKM matrix elements, we take $\vert V_{ub}/\lambda V_{cb}\vert = 0.41$ and  
$\phi_3 =65^\circ,\,85^\circ$.  
We use $|V_{td}|\times10^{3}=8.0,\,9.2$ to get   
$\Delta m_{B_d}^{\rm SM} \sim 0.54,\,0.72\,{\rm ps}^{-1}$,   
respectively, which are close to the experimental value
$\Delta m_{B_d}=0.484\pm0.010\,{\rm ps}^{-1}$ \cite{LEPBOSC}. 
The large uncertainty in $f_{B_d}\hat B_{B_d}^{1/2}$ makes it possible
for $\Delta m_{B_d}^{\rm SM}$ to lie within experimental
range even for large $\phi_3$ .
However, when one considers $\Delta m_{B_s}/\Delta m_{B_d}$, 
the hadronic uncertainty is reduced significantly, i.e.
$\xi_s\equiv f_{B_s}\hat B_{B_s}^{1/2}/f_{B_d}\hat B_{B_d}^{1/2}
=1.16\pm0.05$ from lattice \cite{xis},
and the SM prediction for large $\phi_3$ case is not consistent 
with experiments,
and New Physics would be needed for this case.  

\begin{figure}[t!]
\centerline{{\epsfxsize6 cm \epsffile{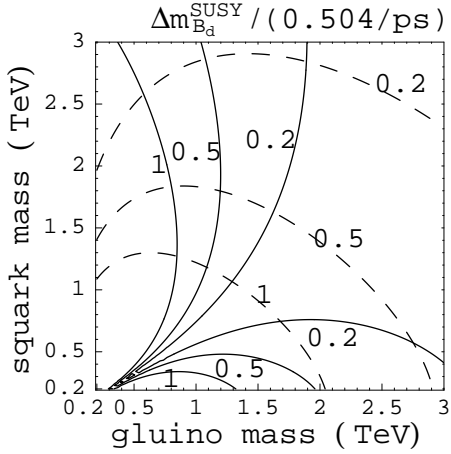}}}  
\smallskip  
\caption {Contribution to $(\Delta m^{\rm SUSY}_{B_d})$ $/$ $(0.504\,{\rm ps}^{-1})$
from gluino box diagrams. 
The solid lines correspond to
$\delta^{13}_{dRR}\sim 0.18$,
while the dashed lines correspond to
$\delta^{13}_{dLL}\delta^{13}_{dRR}\sim (0.18)^4$. 
}  
\label{fig:Bdconst}  
\end{figure}

With the formulas above,
we are ready to discuss the SUSY contributions to 
$B$--$\bar B$ mixing in the AHS models.
In Fig. \ref{fig:Bdconst} we illustrate the
$\widetilde m$--$m_{\tilde g}$ dependence of
$\Delta m^{\rm SUSY}_{B_d}/(0.504\,{\rm ps}^{-1})$ from gluino box diagrams, 
where the denominator is the experimental bound at 2$\sigma$.
The solid lines correspond to contributions from RR-RR mixings
where each squark propagator has one $\delta^{13}_{dRR}\sim 0.18$ insertion.
The dashed lines correspond to contributions from LL-RR mixings
where one squark propagator has 
an insertion with $\delta^{13}_{dLL}\sim 0.18^3$
and the other an insertion with $\delta^{13}_{dRR}\sim 0.18$.

We can tell from Fig. 2 which mixings, LL-RR or RR-RR,
give the dominant contribution in different regions of parameter space.  
For small $m_{\tilde g}$ the RR-RR mixings give tighter constraint,
but for larger $m_{\tilde g}$ the LL-RR mixings is more stringent. 
The parameter space corresponding to $\Delta m^{\rm
SUSY}_{B_d}/(0.504\,{\rm ps}^{-1})\gg1$ is excluded.
Since contributions from other sparticles are sub-dominant 
in most of the parameter space, as we will discuss later, 
they are unlikely to cancel the gluino contribution.

We clearly need TeV range gluino and/or squark masses
to satisfy the $\Delta m_{B_d}$ constraint.
This comes as a result of the large mixings in right-handed sector,
and can be shown by simple arguments.
$C^{\rm W}_1$ in the SM is roughly proportional to 
$(\alpha_{\rm W}/m_{\rm W}^2)^2 m_t^2 (V_{tb} V^*_{td})^2$.
For SUSY contribution assuming $m_{\tilde g}\sim \widetilde m$, 
the $(\alpha_{\rm W}/m^2_{\rm W})^2 m^2_t$ factor is replaced by 
$N_c \alpha^2_S/{\widetilde m^2}$,
and $(V_{tb} V^*_{td})^2\sim\lambda^6$ is replaced by 
$(\delta^{13}_{dRR})^2\sim\lambda^2$ and
$\delta^{13}_{dLL}\delta^{13}_{dRR}\sim \lambda^4$
for RR-RR and LL-RR mixing contributions, respectively. 
Since the SM contribution is already close to 
the experimental observation, one requires 
\begin{equation}
\widetilde m\gtrsim \biggl({\alpha_S\over\alpha_{\rm W}}\biggr) 
                 \sqrt{{N_c\lambda^{4,2}\over \lambda^6}}
                 \biggl({m^2_{\rm W}\over m_t}\biggr)
            \sim 1,\,4\,{\rm TeV},
\end{equation}
from LL-RR and RR-RR mixings, respectively.
Thus the typical scale of superparticles,
$M_{\rm SUSY}$, has to be large due to large squark mixings.
Comparing with Fig. 2, we note that the LL-RR case 
is close to this estimate, 
while $\tilde m$ for the RR-RR case is weaker than the estimate.
This can be traced to the aforementioned cancellation
in the RR-RR case where total cancellation in $\tilde C_1^{\tilde g}$ is
possible for $x_{\tilde g \tilde q}\sim 2.43$.  
Considering RR-RR mixings only, there is a valley in the parameter space
that even light superparticles with masses less than 250 GeV are allowed.

Neutralino box diagrams are induced by the same flavor source as 
the gluino box diagrams.
The neutralino masses could be related to $m_{\tilde g}$ through
a GUT-like relation on the gaugino Majorana masses \cite{Martin:1997ns},
\begin{equation}
\label{GUTrelation}
m_{\tilde g}={\alpha_s\over\alpha_W} M_2,\qquad
M_1={5\over3}{\alpha^\prime\over\alpha_W} M_2.
\end{equation}
The gluino-neutralino box dominates over neutralino-neutralino box.
The neutralino contribution to 
$\Delta m^{\rm SUSY}_{B_d}$ $/$ $(0.504\,{\rm ps}^{-1})$
is less than 10\% of gluino contribution for $\mgl,\msq$ $>$ 500 GeV 
for either RR-RR or LL-RR mixings
with $\tan\beta=2$--50 and $|\mu|=100$--1000 GeV.
For low $\msq$, $|\mu|$ and  large $\tan\beta$, $\mgl$,
its contribution, through $\tilde C^{\tilde \chi^0}_2$, 
can be comparable with the RR-RR mixing induced gluino contribution,
which is suppressed by large $\mgl$.
However, as can be seen from Fig. 2, 
both the RR-RR and LL-RR mixing induced gluino contributions already give 
$\Delta m^{\rm SUSY}_{B_d}$ $/$ $(0.504\,{\rm ps}^{-1})\gg1$
in that region, and we need to fine-tune the relative phase and size of these
mixings to be within experimental bound.

The charged Higgs contributes  
$r$ $\equiv$ $|M^{\rm SUSY}_{12}/M^{\rm SM}_{12}|$
$\sim$ (37, 11, 3)\% for $m_{H^+}$ $=$ (100, 400, 1000) GeV with $\tan\beta=2$,
and $\sim$ (0.3, 0.2, 0.1)\% for $\tan\beta=50$. 
The charged Higgs contribution interferes coherently with the SM amplitude.
The result is consistent with early studies \cite{Branco}.
Without further interference from other SUSY contributions the
$b\rightarrow s\gamma$ branching ratio constrains charged Higgs mass to
be quite large \cite{HouChargedHiggs}.
In this case, the charged Higgs contribution to $B^0$--$\bar B^0$ mixing is small.
It is known that cancellations from other sparticle contributions
may reduce the bound on the charged Higgs mass \cite{Bertolini,ChargedHiggsChargino}.
We will return to this point in Section VI.
On the other hand the chargino gives even smaller contribution,
$r$ $\sim0.1,\,0.5\%$ for TeV range $m_{\tilde g}$ and 
$\mu=\pm1000,\,\pm100$ GeV, respectively.
The main contribution comes from the term with $\hat Y_t^2\delta^{33}_{uRR}$
in $C^{\tilde \chi^-}_1$.
The ratio is reduced by $\sim30\%$ for large $\tan\beta$ mainly due to the 
reduction of $\hat Y^2_t$.
The smallness of chargino contributions is 
in contrast with early studies \cite{Branco}.
In our case there is no large mixing involving third generation in the up 
type sector, and there is no large splitting due to light stop.
In this type of models, the SUSY contribution to $B^0$--$\bar B^0$ mixing
is dominated by gluino exchange.

\begin{figure}[htb]
\centerline{{\epsfxsize6.5 cm \epsffile{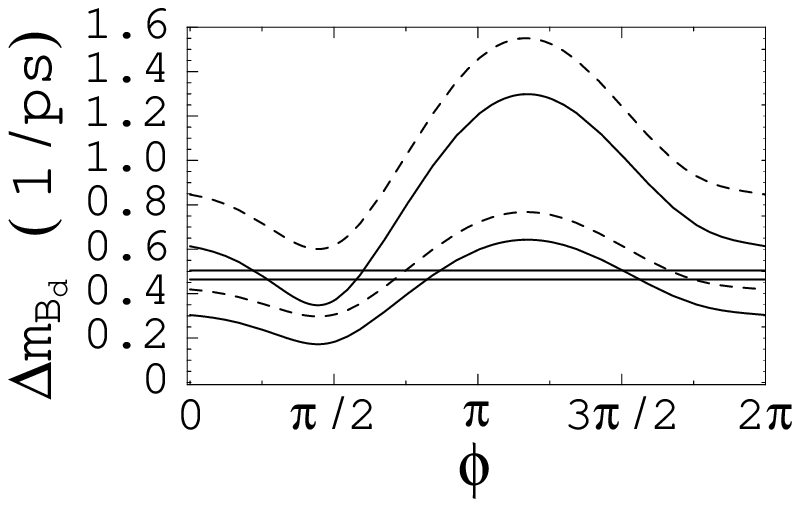}}}  
\centerline{{\epsfxsize6.5 cm \epsffile{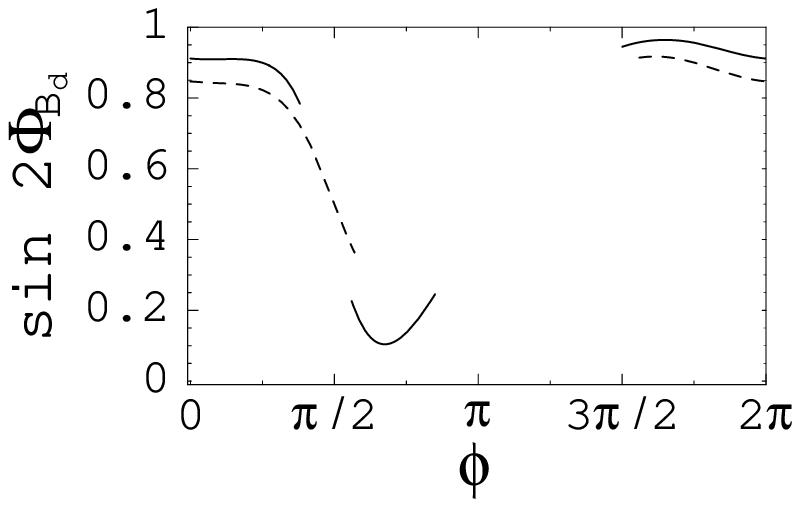}}}  
\smallskip \smallskip 
\caption {  
One sigma range for
(a) $\Delta m_{B_d}$ and (b) $\sin2\Phi_{B_d}$  
 vs. $\phi \equiv \arg \delta_{dRR}^{13}$,   
including both SM and SUSY effects,   
for gluino mass $\mgl =$ 1.5 TeV.   
The solid, short-dashed curves   
correspond to $\phi_3 = 65^\circ$, $85^\circ$
and $\msq= 1.5$ TeV and
$\tan\beta$ is taken as equal to 2.   
The horizontal lines in (a) indicate the $2\sigma$ experimental range.  
}  
\label{fig:Bdlight}  
\end{figure}  

After showing that gluino exchange gives 
dominant contributions to $M_{12}^{\rm SUSY}$,
we turn to explore the interference between 
$M^{\rm SUSY}_{12}$ and  $M^{\rm SM}_{12}$.
We consider the SUSY phase $\phi \equiv \arg \delta_{dRR}^{13}$.
The experimental measurement of $\Phi_{B_d}$ is 
no longer just $\phi_1$ of SM.
For illustration we plot, in Figs. \ref{fig:Bdlight}(a) and (b),  
$\Delta m_{B_d} (\equiv 2\vert M^B_{12}\vert)$ and   
$\sin2\Phi_{B_d}$ vs. $\phi$, respectively,
for 1.5 TeV squark mass and 1.5 TeV gluino mass.
In Fig. \ref{fig:Bdlight}(a), the solid (short-dashed) lines correspond to 
$\phi_3 =65^\circ\,(85^\circ)$.
The upper and lower solid and short-dashed lines denote the 1$\sigma$ boundaries of 
$f_{B_d}\hat B_{B_d}^{1/2}=(230\pm 40)$ MeV.
If RR-RR mixings dominate, the SUSY phase $\phi_{\rm SUSY}\sim 2\phi$,
while if LL-RR mixings dominate, then $\phi_{\rm SUSY}\sim \phi$.
For $\phi_3 =65^\circ\,(85^\circ)$, 
the SUSY model gives $r\sim$22\% (16\%)
from RR-RR mixings, 
and $r\sim$55\% (41\%) from LL-RR mixings.
These are consistent with Fig. 2, since 
$r\sim \Delta m^{\rm SUSY}_{B_d}/(0.504\,{\rm ps}^{-1})$.
The two SUSY contributions interfere constructively (destructively) 
for $\phi\sim\pi\ (0)$.
In the present case,
LL-RR mixings dominate over RR-RR mixings and show a 
$\phi_{\rm SUSY}\sim \phi$ behavior in the graph.

\begin{figure}[t!]
\centerline{{\epsfxsize6.5 cm \epsffile{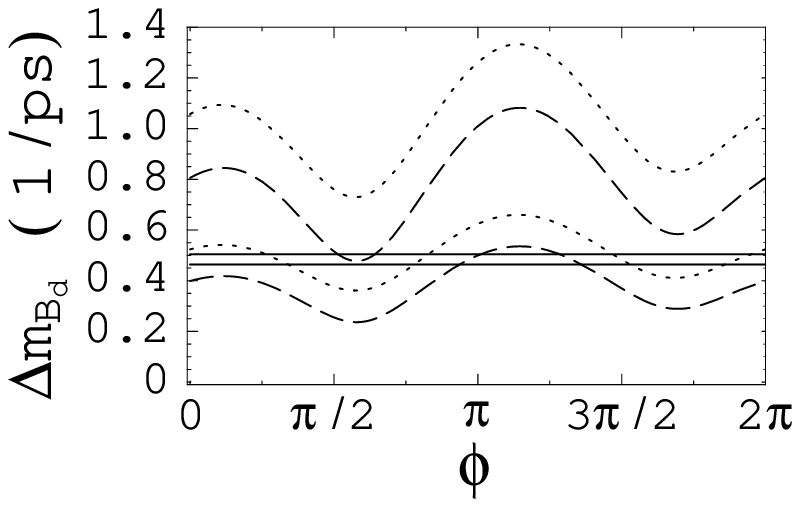}}}  
\centerline{{\epsfxsize6.5 cm \epsffile{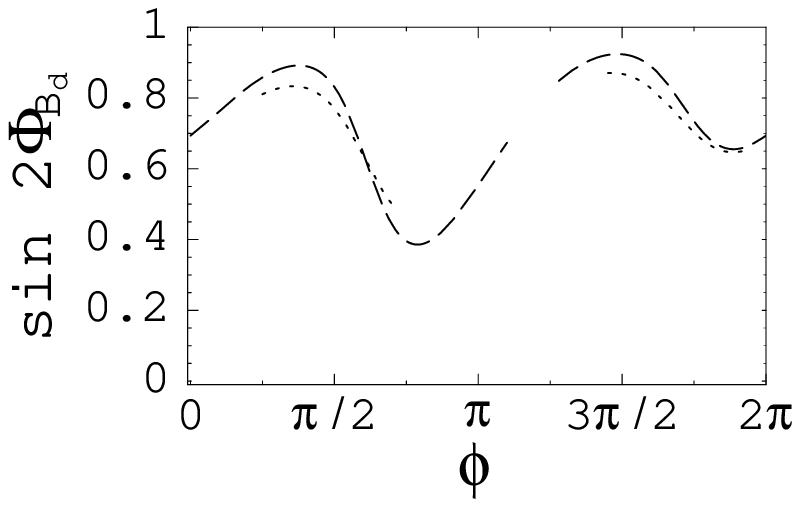}}}  
\smallskip  \smallskip
\caption {  
Same as in Fig. \ref{fig:Bdconst} but for $\msq=$ 3 TeV,
where long-dashed, dotted curves correspond to 
$\phi_3 = 65^\circ$, $85^\circ$.  
}  
\label{fig:Bdheavy}  
\end{figure}

In Fig. \ref{fig:Bdheavy}(a) and (b),  
we show the same physics measurable but with
$\msq=3$ TeV, $\mgl=1.5$ TeV.
For $\phi_3 =65^\circ\,(85^\circ)$, 
the SUSY model contributes $r\sim$29\% (22\%) from RR-RR mixings 
and $r\sim$18\%( 13\%) from LL-RR mixings vs $\Delta m^{\rm SM}_{B_d}$.
These are again consistent with Fig. 2.
The interference pattern is the same as the previous case.
But in the present case, the RR-RR mixing contribution
dominates over LL-RR mixings hence show a 
$\phi_{\rm SUSY}\sim 2\phi$ behavior in Fig. \ref{fig:Bdheavy}(a).
We can see from Fig. \ref{fig:Bdconst} that
RR-RR mixings dominate for the point 
$(m_{\tilde g},\,\widetilde m)=(1.5,\,3)$ TeV.

We see that $\sin2\Phi_{B_d}$ as measured from 
$B_d \to J/\psi K_S$ can range from 0.1--0.95 and 0.4--0.9
as shown in Figs. \ref{fig:Bdlight}(b) and \ref{fig:Bdheavy}(b), respectively.    
These curves are obtained by overlapping various segments from the corresponding 
curves within 1$\sigma$ range of $f_{B_d}{\hat B^{1/2}_{B_d}}$.
For example, in Fig. \ref{fig:Bdlight}(a),
any single line with $\msq=\mgl=1.5$ TeV
that corresponds to a value within 1$\sigma$ of $f_{B_d}\hat B_{B_d}^{1/2}$ 
should lie within the two solid lines.
Each single line only intercepts with the 2$\sigma$ experimental 
range for $\Delta m_{B_d}$ for some region of $\phi$ 
and corresponds to some segment in the solid line of Fig. \ref{fig:Bdlight}(b)
(one can compare with Fig. 1(b) of Ref. \cite{ChuaHou}, 
where $f_{B_d}\hat B_{B_d}^{1/2}=200$ MeV is used).
Taking the uncertainty of $f_{B_d}\hat B_{B_d}^{1/2}$ into account 
enlarges the parameter space considerably.
The fact that these segments from different value of
$f_{B_d}\hat B_{B_d}^{1/2}$ lie on a line and 
not forming a band corresponds to our simplifying 
assumption of using same bag-factor for all ${\cal O}_i$.  
Thus, they can be factored out
without affecting the argument of $M_{12}$ in Eq. (\ref{MB12}).

In SM, we have $\sin2\phi_1 \simeq 0.75$--$0.71$ \cite{Ciuchini:2000de}
for $\phi_3 = 65^\circ$--$85^\circ$. 
The measurements from Belle \cite{Belle} and BaBar \cite{BaBar} in year 2000 
indicated smaller values for $\sin2\phi_1$ vs SM, 
and would correspond to relatively specific 
$\phi \equiv \arg \delta_{dRR}^{13}$ phase values 
(between $\pi/2$ and $\pi$)  
in Figs. \ref{fig:Bdlight}(b) and \ref{fig:Bdheavy}(b).
More recent definitive measurements from 
BaBar \cite{BaBar2} and Belle \cite{Belle2}, 
in year 2001, give $\sin2\phi_1$ values close to 1! 
The average of BaBar and Belle 2001 values is given in 
Eq.~(\ref{sin2phi1}). 
It is rather intriguing that this range corresponds to
the main parameter space allowed by $B_d$ mixing,
suggesting that SUSY contributions could be comparable to SM.
In particular, if the Belle 2001 central value of
$\sin2\phi_1 = 0.99$ holds up, it would imply that
$\phi \equiv \arg \delta_{dRR}^{13}$ is between
$3\pi/2$ and $2\pi$. 
The lighter squark mass case of Figs. \ref{fig:Bdlight}
is preferred, but the heavier squark mass case of 
Figs. \ref{fig:Bdheavy} is also possible.

To conclude this section, we note that one requires heavy squark and gluino
masses to satisfy $\Delta m_{B_d}$ bound.
While the direct search of such heavy superparticles become less promising, 
these particles can show their effect in
$B_d$--$\bar B_d$ mixing phases, even with TeV masses.

\subsection{Brief Discussion on $B_s$ Mixing and $CP$ Phase}

As shown in previous section, one could have
$s$ flavor decoupled and the above discussion is applicable,
and is being tested right now.
Alternatively, and mutually exclusive to the above case,
it could be the $d$ flavor that is decoupled,
and the $B_d$ system would be SM-like, which may still
turn out to be the case in 2002.
If so, $B_s$ mixing may be the place where
SUSY AHS effects show up.

With the CKM like relation $\delta^{13}_{dRR}/\delta^{23}_{dRR}\sim
V_{td}/V_{ts}\sim\lambda$ from Eqs. (\ref{MRR}), (\ref{delta}),
the $B_s$ mixing case is rather similar to 
the discussion of $B_d$ mixing, 
so long that the mass insertion approximation can be used.
One can just scale up from previous discussion.
The gluino contribution to $B^0_s$--$\bar B^0_s$ mixing is 
discussed in Ref. \cite{Barenboim} and \cite{sb},
where in the latter work the mass insertion approximation is relaxed.
Since $\tilde s_R$--$\tilde b_R$ mixing $\sim 1$ in SUSY AHS model,
one in principle could have a relatively light ``strange-beauty" squark,
which, unlike the heavy SUSY scale that is the focus of this
paper, can impact on direct search.
We will discuss this case later in Sec. VII.

The $B_s$ mixing phase may not be vanishingly small as in SM, 
and can be searched for at the Tevatron collider in a matter of years.
The most interesting situation would be to find (soon!)
$\Delta m_{B_s}$ not far above SM expectation,
but with large $\sin2\Phi_{B_s}$.

\section{Constraints from $K^0$--$\bar K^0$ mixing}   
 
As shown in Eqs. (\ref{Mq}) and (\ref{MRR}),
AHS models not only give large mixing in RR sector involving third generation
down squark, they also give large mixing in 1-2 generations.
It is well known that $\Delta m_K$ is much smaller   
than $\Delta m_{B_d}$ hence offers a much stronger constraint,  
while $\varepsilon_K$ is even tighter.  
They make $\delta^{12}_{dLL,RR} \sim \lambda$ impossible to sustain  
even with $\widetilde m$, $m_{\tilde g} \gtrsim$ TeV
\cite{Silvestrini,kaon}.
The formulas for kaon mixing are similar to that for $B_q$--$\bar B_q$ mixing,
with only some modifications needed.
For charged Higgs exchange diagrams, Eq. (\ref{CH}) now becomes,
\begin{eqnarray}  
C^{H^-}_1&=&{\alpha_{\rm W}^2\over 8 m_W^2} V_{is} V_{js} V^*_{id} V^*_{jd}   
          [x_{i\rm W}\, x_{jH} \cot^4 \beta {1\over 4} G(x_{iH},x_{jH})  
\nonumber \\             
&&\quad+2 x_{i\rm W} x_{j\rm W} \cot^2 \beta   
 (F^\prime(x_{i\rm W},x_{j\rm W},x_{H\rm W})
\nonumber \\
&&\qquad+{1\over 4}G^\prime(x_{i\rm W},x_{j\rm W},x_{H\rm W})],  
\end{eqnarray}
where $i,j$ are generation indices of up type quarks and summed over.
Other terms are neglected due to the smallness of quark masses $m_{s,d}$.
Chargino contributions are modified by changing
$V^*_{lq}V_{mb} \delta^{lm}_{uLL}$ in Eq. (\ref{CX}) 
to $V^*_{ld}V_{ms} \delta^{lm}_{uLL}$
and neglecting other terms.
Gluino and neutralino contributions are modified by
changing $\delta^{q3}$ in Eqs. (\ref{Cg}), (\ref{CgX0}), (\ref{CX0})
to $\delta^{12}$ and neglecting all $H_{L,R}$ terms.
The QCD running formula is also modified accordingly.

\begin{table}[b!]  
\label{Kaon}  
%
\begin{center} 
\begin{tabular}{cccc}  
$x$  
   &$\sqrt{|{\rm Re}\,(\delta^{12}_{dLL})^2|}$ 
           &$\sqrt{|{\rm Re}\, \delta^{12}_{dLL} \delta^{12}_{dRR}|}
            $
                    &$\sqrt{|{\rm Re}\,(\delta^{12}_{dRR})^2|}$ \qquad
\\ \hline 
0.3
   &$\ 8.9\times10^{-2}$
           &$\ 4.3\times10^{-3}$
                    &$\ 9.3\times10^{-2}$  
\\ 
1
   &$\ 2.3\times10^{-1}$ 
           &$\ 4.9\times10^{-3}$
                    &$\ 2.0\times10^{-1}$  
\\ 
4
   &$\ 2.5\times10^{-1}$
           &$\ 7.0\times10^{-3}$ 
                    &$\ 4.3\times10^{-2}$
\\ \hline
   &$\sqrt{|{\rm Im}\,(\delta^{12}_{dLL})^2|}$  
           &$\sqrt{|{\rm Im}\, \delta^{12}_{dLL} \delta^{12}_{dRR}|}$  
                    &$\sqrt{|{\rm Im}\,(\delta^{12}_{dRR})^2|}$  
\\ \hline
0.3
   &$\ 7.1\times10^{-3}$
           &$\ 3.4\times10^{-4}$
                    &$\ 7.4\times10^{-3}$  
\\ 
1
   &$\ 1.8\times10^{-2}$
           &$\ 3.9\times10^{-4}$
                    &$\ 1.6\times10^{-2}$  
\\ 
4
   &$\ 2.0\times10^{-2}$
           &$\ 5.6\times 10^{-4}$
                    &$\ 3.4\times 10^{-2}$  
\end{tabular}  
{\caption{Limits on $\sqrt{|{\rm Re}\,\delta_d^{12}\,\delta_d^{12}}|$  
and $\sqrt{|{\rm Im}\,\delta_d^{12}\,\delta_d^{12}}|$  
for squark mass $\widetilde m=1.5$ TeV and  
for different $x=m^2_{\tilde g}/{\widetilde m}^2$, 
including leading order QCD  corrections.  
The constraints are from   
$\Delta m_K^{\rm SUSY}<3.521\times 10^{-12}$ MeV and  
$|\varepsilon_K^{\rm SUSY}|<2.268\times10^{-3}$.  
}}  
\end{center} 
\end{table}  

The charged Higgs contributions are in general small.
For gluino and neutralino contributions we show in Table~II 
the limits on $\sqrt{|{\rm Re}\,\delta^{12}_{AB}\delta^{12}_{CD}|}$ and  
$\sqrt{|{\rm Im}\,\delta^{12}_{AB}\delta^{12}_{CD}|}$
from $(\Delta m^{\rm SUSY}_K)<3.521\times 10^{-12}$ MeV
for $\widetilde m=1.5$ TeV, where $A,B,C,D=L,R$.  
The constraints from $\varepsilon_K$ can also be estimated by using  
\begin{equation}
\label{epsilon}  
|\varepsilon_K|={|{\rm Im}\,M_{12}|\over {\sqrt 2}\Delta m_K}<2.268\times10^{-3}.  
\end{equation}   
For different values of $\widetilde m$, the limits can be 
roughly obtained by multiplying   
a factor $\widetilde m/(1.5\, {\rm TeV})$.    
We take the hadronic scale in Eq. (\ref{etas}) to be $\sim$ GeV and   
$\alpha_s(M_{\rm Z})=0.1185$.   
We can reproduce the results of Ref. \cite{Bagger} by using 
$\alpha_s(\mu)\sim 1$  
and $\widetilde m=500$ GeV and by considering gluino contributions only.
The QCD effects enhance the SUSY contributions   
by a few times for contributions arising from  
$\delta_{dLL}\,\delta_{dRR}$, as one can see from $\eta_4\sim 6$.    
Since the most severe constraint is on $\delta_{dLL} \delta_{dRR}$,
the QCD effects make it more stringent \cite{Bagger}.   
From Eq.~(\ref{epsilon}), 
$\varepsilon_K$ gives even more stringent constraint than $\Delta m_K$. 
However, the constraint becomes less severe if 
the phases of $\delta$s are small (of order $0.01$).  
Together with constraint from electric dipole moment of neutron,
one may be led to the idea of approximate CP (see for example \cite{approxCP}).

From the above discussion, 
it is clear that $\delta^{12}_{dLL},\,\delta^{12}_{dRR}\sim\lambda$ 
cannot be sustained. 
One has to invoke QSA as discussed in Sec. II to
impose appropriate ``texture zeros".
We see that the values given in Table I are all well below   
the limits from $\Delta m_K$ and $\varepsilon$ constraints,
Table II, even with $\cal O$(1) phases.
The approximate CP assumption can be relaxed.

\begin{figure}[t!]  
\centerline{\DESepsf(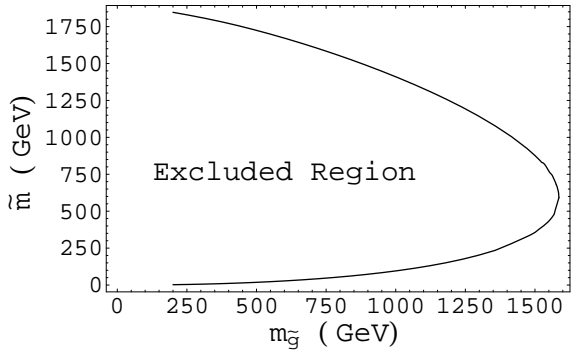 width 7.5cm)}  
\smallskip  
\caption{  
Limit on $\widetilde m$ vs $m_{\tilde g}$
from chargino contributions to $\Delta m_K$
arising from $\delta^{12}_{uLL}\sim\lambda$.   
}  
\label{fig:kaon}
\end{figure}

As noted in Sec. II,
QSA will induce $\delta^{12}_{uLL} \sim \lambda$
by shifting the source of the Cabibbo angle to the up-type sector.
The full strength $\delta^{12}_{uLL}$
can contribute to kaon mixing via chargino diagrams.
In Fig. \ref{fig:kaon}, we show the parameter space constrained by
$\Delta m_K$.
We use the GUT relation on gaugino mass as given in 
Eq. (\ref{GUTrelation}) for sake of simplicity and definiteness. 
The horizontal axis can be converted to wino mass
by multiplying $\mgl$ by $\sim0.4$.
For the $\varepsilon_K$ constraint, 
we would need arg$(\delta_u)$ to be less than 0.1.
We see from Fig. \ref{fig:kaon} that 
the kaon mixing constraint also points to 
TeV scale gluino and squarks. 
We stress that this is a generic feature of QSA and 
has nothing to do with the choice of retaining $M^{3i}_d$ or not.
Keeping $M^{3i}_d$ leads to interesting low energy physics,
even with TeV scale particles as a result of kaon mixing constraints.

We note that it is possible for 
the chargino diagrams to interfere destructively with 
LL (or RR) mixing induced gluino and neutralino contributions, 
and one can satisfy the kaon constraint with a lower mass scale. 
However, since these correspond to different set of parameters, 
it is unlikely for the interference to be destructive in general.

\section{Implications for $D^0$--$\bar D^0$ Mixing}

The experimental situation for $D^0$--$\bar D^0$ mixing
is rather volatile at the present time.
A search by the CLEO Collaboration gives
$1/2 x_D^{\prime2}<0.041\%$ and  $-5.8\%<y_D^\prime<1.0\%$ \cite{xD1},
where $x_D^\prime$ and $y_D^\prime$ are defined in Eq. (\ref{xpyp}).   
CLEO further adopted $\delta_D \simeq 0$ from model arguments  
to reach a more stringent bound of $x_D\simeq x_D^\prime < 2.9\%$.  
If $\delta_D \neq 0$ \cite{strongphase}, however,   
the preferred negative value of $y_D^\prime$ may in fact 
be hinting at $x_D\sim$  the few \% level.

\begin{figure}[t!]
\centerline{\DESepsf(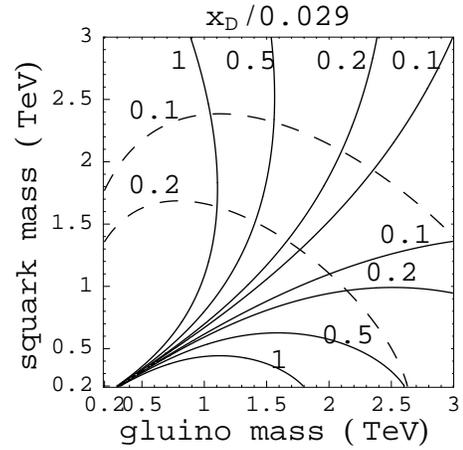 width 6cm)}
\smallskip\smallskip
\caption{Contribution to $x_D/2.9\%$ from gluino box diagrams.
The solid lines correspond to $\delta^{12}_{uLL}\sim 0.18$,
while the dashed lines correspond to 
$\delta^{12}_{uLL}\delta^{12}_{uRR}\sim(0.18)^{5.5}$.
}
\label{fig:Dmixconst}
\end{figure}


Another approach is to compare $D^0 \to K^- \pi^+$ and $K^- K^+$ decays
and measure the lifetime difference between $CP$ even and odd final states.
The current world average from Belle\cite{Belleycp},
BaBar\cite{BaBarycp}, CLEO\cite{CLEOycp}, E791\cite{E791} and
FOCUS\cite{xD2} Collaborations is $(1.1\pm0.99)$\%\cite{Ligeti}. 
Since this is consistent with zero
and does not support the nonzero claim by FOCUS, 
we shall take the more stringent constraint of $x_D<2.9\%$ 
from CLEO in the following. 
What we find intriguing, however, 
is that   $\delta^{12}_{uLL} \sim \lambda$ 
with $\widetilde m$, $m_{\tilde g} \sim $ TeV
brings $x_D$ right into the ball-park of the \% level!
Furthermore,
this can be probed in detail in the next few years
at the B factories, and in the longer run,
by hadron collider detectors.

\begin{figure}[b!]  
\centerline{{\epsfxsize7.5cm \epsffile{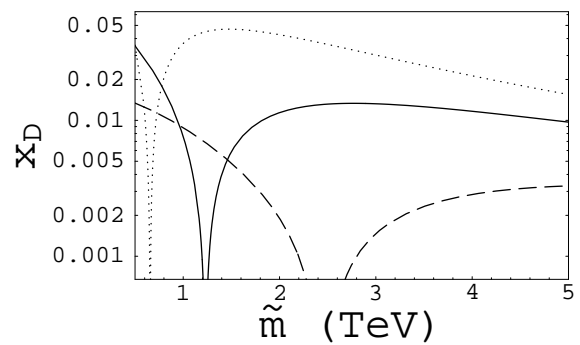}} }  
\smallskip  
\caption{Gluino contribution to $x_D$ vs. $\widetilde m$.
Dotted, solid and dashed lines are for   
$m_{\tilde g} = 0.8$, 1.5 and 3 TeV, respectively.  
}  
\label{fig:Dmix}  
\end{figure}

We consider gluino and neutralino exchange diagrams 
induced by up-squark mixing with $\mu=1$ TeV and $\tan\beta=2$ with
formulas similar to $B$ mixing. 
The dependence on $\tan\beta$ is weak.
The SUSY contribution from gluino box to $x_D/0.029$
is illustrated in Fig.~\ref{fig:Dmixconst} in the $\mgl$--$\msq$ plane.
The solid lines correspond to $\delta^{12}_{uLL}\sim 0.18$,
while the dashed lines correspond to 
$\delta^{12}_{uLL}\delta^{12}_{uRR}\sim(0.18)^{5.5}$.
It is clear that LL-LL induced gluino box diagrams dominate. 
In Fig.~\ref{fig:Dmix} we illustrate $x_D$ vs. $\widetilde m$ 
for $m_{\tilde g}=0.8,\,1.5,\,3$ TeV, respectively.  
As in the $B$ mixing case, there is a narrow valley from $\delta_{uLL}$ 
induced gluino contributions around $m_{\tilde g}^2/{\widetilde m}^2\sim2.43$
when $C^{\tilde g}_1$ of Eq. (\ref{Cg}) vanishes.  
This could make the parameter space from Fig. \ref{fig:Dmixconst} too restrictive.
However, the actual zeros in Fig. \ref{fig:Dmix},
occur at slightly shifted mass ratios, 
reflecting a cancellation between various contributions from $\delta_{uLL}$ and 
$\delta_{uLL}\delta_{uRR}$ when they have a common phase.
Although $\varepsilon_K$ constrains $\arg(\delta^{12}_{uLL})$ to be less
than  0.1, the phase of $\delta^{12}_{uRR}$ is not constrainted
since $\delta^{12}_{uRR}$ is by itself small.
In general, the SUSY phase $\delta^{12}_{uRR}$ does not have to vanish,   
and having phase in common with $\delta_{uLL}$ is not likely.
Thus, the deep valley would in general be filled,
but the figure illustrates the adjustability of $x_D$.   
It also gives an explicit example where   
detectable $D^0$ mixing would likely \cite{Wolf_D}   
carry a $CP$ violating phase.

To conclude this section, we note that AHS with QSA is 
known to produce large $D$ meson mixings.
The stringent upper bound on $x_D$ seem to provide severe constraint for
QSA models \cite{Nir93,Nelson}.
This is more or less true when squarks and gluino are 
as light as a few hundred GeV. 
However, as a result of the large mixing in squark sector in our case,
the proximity of $\Delta m_{B_d}$ to SM expectation
leads to squarks at TeV scale, 
and $\sin2\phi_1$ may be affected in an interesting way. 
It is intersting that the scale determined from
this leads to a $D^0$ meson mixing close to experimental hints.
We eagerly await the experimental situation to clear up,
i.e. whether the CLEO hint is due to $\Delta\Gamma_D$ \cite{Ligeti} or
$\Delta m_D$.

\begin{figure}[b!]
\centerline{\DESepsf(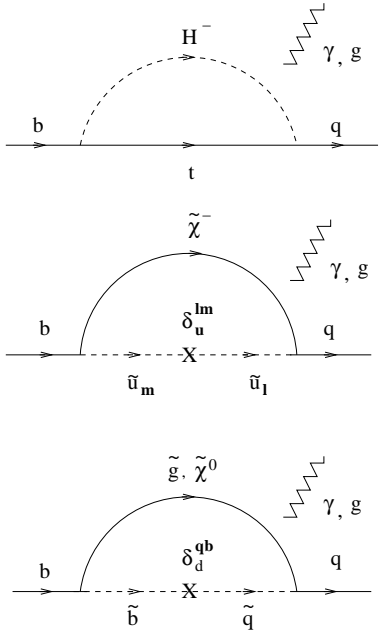 width 5cm)}
\smallskip
\caption{SUSY penguin diagrams for $b\to q\gamma,q g$ processes.   
}
\label{b2qph}
\end{figure}

\section{Radiative $B$ Decays}

The effective Hamiltonian     
for $b\rightarrow q\gamma,\,qg$ transitions, 
where $q=d$ or $s$, is given by
\begin{eqnarray}  
H_{{\rm eff.}} &=&-{G_{F}\over \sqrt{2}}{m_b\over 4\pi^2}
V_{tb}V_{tq}^{*}\bar{q}\left\{\,e\left[ C_{7\gamma}\,R+  
C_{7\gamma}^{\prime}\,L\right] F^{\mu \nu} \right.
\nonumber \\  
&&\quad\left.+\,g\left[ C_{8g}\,R+  
C_{8g}^{\prime }\,L\right]  
T^{a}G_{a}^{\mu \nu }\right\}\sigma _{\mu \nu }b,
\end{eqnarray}  
where we have neglected $m_{q}$,   
$C_{7,8}=C_{7,8}^{{\rm SM}}+C_{7,8}^{{\rm New}}$   
are the sum of SM and New Physics contributions,   
while $C_{7,8}^{\prime }$ come purely from New Physics.   
We are particularly interested in the case where $C^\prime_{7\gamma,8g}$
are large.
The effects from the SUSY contributions are given by  
\begin{eqnarray}  
C_{7\gamma}^{(\prime)\rm New} &=&   
C^{(\prime)}_{7\gamma,H^-}+C^{(\prime)}_{7\gamma,\tilde g}
+C^{(\prime)}_{7\gamma,\tilde \chi^-}  
+C^{(\prime)}_{7\gamma,\tilde \chi^0},  
\\  
C_{8g}^{(\prime)\rm New} &=&   
C^{(\prime)}_{8g,H^-}+C^{(\prime)}_{8g,\tilde g}+C^{(\prime)}_{8g,\tilde \chi^-}  
+C^{(\prime)}_{8g,\tilde \chi^0}.
\end{eqnarray}
The Feynman diagrams are shown in Fig. \ref{b2qph}.

\subsection{Formulas}

\noindent\underline{Charged Higgs Exchange}:  
\begin{eqnarray}  
C_{7\gamma,H^-} &=& -{x_{tH}\over 2}   
\biggl\{\cot^2\beta\left[Q_u F_1 (x_{tH})+F_2 (x_{tH})\right]  
\nonumber \\      
&&\quad\qquad+\left[Q_u F_3 (x_{tH})+F_4 (x_{tH})\right] \biggr\},  
\label{C7ChargedHiggs}
\\  
C_{8g,H^-} &=& -{x_{tH}\over 2} \biggl[  
\cot^2\beta F_1 (x_{tH})+F_3 (x_{tH})\biggr],  
\end{eqnarray}  
where $F_i(x)$ are loop functions and the explicit
expressions can be found in Ref.~\cite{Bertolini}.
   
\vspace{0.5cm}
\noindent\underline{Gluino Exchange}:  
\begin{eqnarray}  
\label{c7New}
&&C_{7\gamma,\tilde g}=   
\frac{\pi \alpha _{s}}{\sqrt{2}{G}_{F}V_{tb}V_{td}^{*}}  
 \frac{Q_{d}2\,C_{2}(R)}{{\tilde m}^2}
\nonumber \\  
&&\qquad\qquad\quad\times\biggl\{ \delta^{13}_{dLL}\, g_{2}(x_{\tilde g \tilde q})
-\frac{m_{\widetilde{g}}}{m_{b}}\,  
         \delta^{13}_{dLR}\, g_{4}(x_{\tilde g \tilde q})  
 \biggr\},\\  
&&C_{8g,\tilde g} =  
 \frac{\pi \alpha_{s}}{\sqrt{2}{G}_{F} {\tilde m}^2 V_{tb}V_{td}^{*}} 
\nonumber \\
&&\qquad \times\biggl\{ \delta^{13}_{dLL}   
 \biggl[[2C_2(R)-C_2(G)]\, g_2(x_{\tilde g \tilde q})  
 -C_2(G)\, g_1(x_{\tilde g \tilde q})\biggr]  
\nonumber \\  
  &&\quad\qquad+\frac{m_{\widetilde g}}{m_b} \,  
    \delta^{13}_{dLR}  
  \biggl[[C_2(G)-2C_2(R)]\, g_4(x_{\tilde g \tilde q})
\nonumber \\
&&\quad\qquad\qquad\qquad\qquad\qquad\qquad
  +C_2(G)\, g_3(x_{\tilde g \tilde q})  
  \biggr] \biggr\},   
\label{c8New}  
\end{eqnarray}  
where $Q_{d}$ is the down quark electric charge,  
$C_{2}(G)=N=3$ and $C_{2}(R)=(N^{2}-1)/(2N)=4/3$ are Casimirs,  
and the functions  
$g_i(x)=-{d/dx} [x\,F_i(x)]$, i.e.   
\begin{eqnarray}  
g_{1}(x) &=&\frac{1+9x-9x^2-x^3+6x(1+x)\,\ln x}{6\,(x-1)^{5}},  
 \nonumber\\  
g_{2}(x) &=&\frac{1-9x-9x^2+17x^3-6x^2(3+x)\,\ln x}{12\,(x-1)^{5}},  
 \nonumber\\  
g_{3}(x) &=&\frac{5-4x-x^2+(2+4x)\,\ln x}{2\,(x-1)^{4}}, 
\nonumber\\  
g_{4}(x) &=&\frac{-1-4x+5x^2-2x(2+x)\,\ln x}{2\,(x-1)^{4}}.  
\end{eqnarray}  
The chirality partners $C_{7\gamma,8g}^{\prime }$  
are obtained by interchanging $L$ and $R$ in the $\delta$'s.  
We see that $\delta_{LL}$ and $\delta_{LR}$   
contribute to $C_{7\gamma,8g}$,  
while $\delta_{RR}$ and $\delta_{RL}$ contribute to   
$C^\prime_{7\gamma,8g}$.  
There is an enhancement factor   
$m_{\tilde g}/m_b$ that comes with $\delta_{LR,RL}$.  
The factor $m_b$ is from normalizing with respect to the SM result,
and hence the smallness of the $b$ quark mass with respect  
to the gluino mass is the origin of this enhancement.
Such enhancement was noted in our earlier study~\cite{b2sp}   
of $\tilde s$-$\tilde b$ mixings  
where $\tilde d$ sector was decoupled completely.  
It has also been invoked to generate $\varepsilon^\prime/\varepsilon$  
via an analogous $\delta^{12}_{LR}$ term~\cite{Masiero,murayama}  
under a horizontal U(2) (hence non-Abelian) symmetry model.  
The mechanism is generic and has been discussed in Ref. \cite{Fujikawa},  
but SUSY with LR squark mixings gives a beautiful example.

\vspace{0.5cm}
\noindent\underline{Chargino Exchange}:  
\begin{eqnarray}  
&&C_{7\gamma,\tilde\chi^-}=  
{m^2_w\over \widetilde m^2 V_{tb} V^*_{tq}}
\nonumber \\
&&\times  
\biggl\{ \biggl[{\cal V}_{j1} {\cal V}^*_{j1} V^*_{lq} V_{mb} \delta^{lm}_{uLL}  
-{\cal V}_{j1} {\cal V}^*_{j2}\hat Y_t V^*_{lq}V_{tb}  \delta^{lt}_{uLR}
\nonumber \\  
&&\quad
-{\cal V}_{j2} {\cal V}^*_{j1}\hat Y_t V^*_{tq} V_{mb} \delta^{tm}_{uRL}  
+{\cal V}_{j2} {\cal V}^*_{j2} \hat Y_t^2 V^*_{tq} V_{tb} \delta^{tt}_{uRR}\biggr]
\nonumber \\  
&&\times
\biggl[g_1 (x_{\tilde\chi^-_j\tilde q})+Q_u g_2 (x_{\tilde\chi^-_j\tilde q})\biggr]  
-{m_{\tilde\chi^-_j}\over m_b} \biggl[{\cal V}_{j1} {\cal U}_{j2} \hat Y_b   
 V^*_{lq} V_{mb} \delta^{lm}_{uLL}
\nonumber \\
&&\quad
-{\cal V}_{j2}{\cal U}_{j2}\hat Y_t \hat Y_b V^*_{tq} V_{mb} \delta^{tm}_{uRL}\biggr]
\biggl[g_3 (x_{\tilde\chi^-_j\tilde q})+Q_u g_4 (x_{\tilde\chi^-_j\tilde q})\biggr]  
\biggr\},  
\nonumber\\
\label{C7chargino}
\\  
&&C^\prime_{7\gamma,\tilde\chi^-}=  
\hat Y_q\, {m^2_w\over \widetilde m^2 V_{tb} V^*_{tq}}
\nonumber \\  
&&\times
\biggl\{ {\cal U}_{j2}{\cal U}^*_{j2}\hat Y_b V^*_{lq}V_{mb}\delta^{lm}_{uLL}  
\biggl[g_1 (x_{\tilde\chi^-_j\tilde q})+Q_u g_2 (x_{\tilde\chi^-_j\tilde q})\biggr]  
\nonumber\\  
&&\quad  
+{m_{\tilde\chi^-_j}\over m_b} 
\biggl[g_3 (x_{\tilde\chi^-_j\tilde q})+Q_u g_4 (x_{\tilde\chi^-_j\tilde q})\biggr]
\nonumber \\
&&\times
\biggl[{\cal V}^*_{j2} {\cal U}^*_{j2} \hat Y_t   
 V^*_{lq} V_{tb} \delta^{lt}_{uLR}
-{\cal U}^*_{j2}{\cal V}^*_{j1} V^*_{lq} V_{mb} \delta^{lm}_{uLL}\biggr] 
\biggr\},  
\end{eqnarray}
where as before we sum over $l,\,m$ for three generations and $j$ for
two chargino mass eigenstates.  
$C_{8g,\tilde\chi^-}^{(\prime)}$ can be obtained by dropping $g_{1,3}$ from
above equations and replacing $Q_u$ with 1.
It is clear from these equations that $C^\prime_{7\gamma,\tilde\chi^-}$ is
suppressed by $\hat Y_q$.

\vspace{0.5cm}
\noindent\underline{Neutralino Exchange}: 
\begin{eqnarray}
&&C_{7\gamma,\tilde\chi^0}={Q_d m^2_w \over \widetilde m^2 V_{tb} V^*_{tq}}  
\biggl\{ 2 G^{*j}_{qL} G^j_{bL} \delta^{i3}_{LL} g_2 (x_{\tilde\chi^0_j\tilde q})  
\nonumber \\  
&&\quad  
+{m_{\tilde\chi^0_j}\over m_b} \biggl[-2 G^{*j}_{qL} G^j_{bR} \delta^{i3}_{dLR}  
 +\sqrt 2  G^{*j}_{qL} H^j_{bL} \delta^{i3}_{dLL}\biggr]  
  g_4 (x_{\tilde\chi^0_j\tilde q})  
\biggr\},
\nonumber \\
\label{c7X0}   
\end{eqnarray}  
where $j$ is summed over four neutralino mass eigenstates,
and $C^{(\prime)}_{8g,\tilde\chi^0}$ can be obtained by   
replacing $Q_d\rightarrow 1$ in the above equation.
We neglect terms with $H^*_{qL,R}$ and some LR mixing terms when there is
no chiral enhancement. 
Similar to the gluino case, the chiral partners $C^\prime$ are obtained by 
taking a conjugation in the chirality $L\leftrightarrow R$ and noting that
$G_{L,R}\leftrightarrow -G_{R,L}$.
  
When running down to the B decay scale $\mu \approx m_{b}$,  
the leading order Wilson coefficients $C_{i}^{(\prime )}$   
are given by \cite{c7c8mix},   
\begin{eqnarray}  
C_{7\gamma}(\mu&=&m_{b})=-0.31+\eta_7\,C_{7\gamma}^{{\rm New}}(M_{\rm SUSY})
\nonumber \\  
&&\qquad\qquad +{8\over3}(\eta_8-\eta_7)\,C_{8g}^{{\rm New}}(M_{\rm SUSY}),  
 \nonumber\\  
C_{8g}(\mu&=&m_{b})=-0.15+\eta_8\,C_{8g}^{{\rm New}}(M_{\rm SUSY}),  
\end{eqnarray}  
where,  
\begin{eqnarray}  
\eta_7&=&\biggl({\alpha_s(M_{\rm SUSY})\over\alpha_s(m_t)}\biggr)^{16/21}\,  
       \biggl({\alpha_s(m_t)\over\alpha_s(m_b)}\biggr)^{16/23},  
\nonumber \\  
\eta_8&=&\biggl({\alpha_s(M_{\rm SUSY})\over\alpha_s(m_t)}\biggr)^{14/21}\,  
       \biggl({\alpha_s(m_t)\over\alpha_s(m_b)}\biggr)^{14/23},  
\end{eqnarray}  
while for opposite chirality, which receives no SM contribution,   
one simply replaces $C^{{\rm New}}$ by $C^{\prime }$   
and set the constant terms to zero.

\subsection{Phenomenological Impact}

It is well known that $B\to X_s \gamma$ is a severe constraint on New Physics. 
The current experimental results are 
$Br(B\to X_s\gamma)=
(2.85\pm0.35\pm0.22)\times10^{-4}\cite{CLEOsgamma},
(3.37\pm0.53\pm0.42{+0.50\atop-0.54}|_{\rm model})\times 10^{-4}\cite{Bellesgamma},
(3.11\pm0.82\pm0.72)\times 10^{-4}\cite{ALEPHsgamma}$,
from CLEO, Belle and ALEPH, respectively.
It is known that the charged Higgs contribution 
interferes constructively \cite{HouChargedHiggs}
with SM contribution $C^{\rm SM}_{7\gamma}(m_b)=-0.31$. 
By using Eq.(\ref{C7ChargedHiggs}), we have, 
\begin{equation}
{C_{7\gamma,H^-}\over C^{\rm SM}_{7\gamma}}=1+(35\%,22\%,15\%,11\%,9\%),
\label{chargedHiggs}
\end{equation}
for $m_{H^-}=(400, 600, 800, 1000, 1200)$ GeV.
The rate can get enhanced by $\sim$ 80\%--20\%.
The result is insensitive to $\tan\beta$ within 2--50, since the term without 
$\cot\beta$ in Eq. (\ref{C7ChargedHiggs}) is dominant\cite{HouChargedHiggs}. 
If we require that the deviation from the SM rate to be less than 20\%,
which is close to experimental error, we need $m_{H^+}\geq 1.2$ TeV,
or we may need cancellations from other particles
\cite{Bertolini}.
The chargino contribution may partially cancel 
the charged Higgs contribution \cite{ChargedHiggsChargino}.
This mechanism is still operative even if we have 
$m_{\tilde g},\widetilde m=1.5$ TeV.
\begin{eqnarray}
{C_{7\gamma,\tilde \chi^-}\over C^{\rm SM}_{7\gamma}}
&=&1\pm
\biggl\{
{\begin{array}{c}
13\;(0.6)\\
17\;(0.9)\\
12\;(0.6)
\end{array}}\biggr\}\%
{\delta^{33}_{uRL}\over m_t/\msq}
\nonumber \\
&&\quad
+\biggl\{
{\begin{array}{c}
18\;(0.5)\\
8\;(0.05)\\
3\;(0.08)
\end{array}}\biggr\}\%
{V^*_{ls} V_{mb} \delta_{uLL}^{lm}\over\lambda^2}
\nonumber \\
&&\quad  -
\biggl\{
{\begin{array}{c}
0.3\;(0.4)\\
0.2\;(0.2)\\
0.1\;(0.1)
\end{array}}\biggr\}\%{V^*_{ts} V_{tb}\delta_{uRR}^{33}\over \lambda^2},
\label{chargino}
\end{eqnarray}
for $\mu=\pm(100,500,1000)$ GeV, and $\tan\beta=50\;(2)$.
The $\delta^{33}_{uRL}$ term comes from chiral enhancement 
(with $m_{\tilde\chi^-_j}/m_b$ factor).
The sign of the coefficient in front of $\delta^{33}_{uRL}$
is the same as the sign of $\mu$.
The coefficient for $|\mu|=500$ GeV is greater than $|\mu|=100$ GeV,
because of chiral enhancement.
The coefficient will drop below 0.1 for $|\mu|\geq1.2$ TeV.
LR and RL mixings without chiral enhancement are negligible,
their contributions being only about $10^{-5,-6}$ of $C^{\rm SM}_{7\gamma}$.
The term with $V^*_{ls} V_{mb} \delta_{uLL}^{lm}$ is also
from  the chiral enhancement term, while the last term is not chirally
enhanced. Due to the smallness of $\hat Y_s$,
$|C^\prime_{7\gamma,\tilde\chi^-}/C^{\rm SM}_{7\gamma}|$ is below 1\%
for expected mixing angles .
For $\tan\beta=2$, terms are suppressed by the 
$\hat Y_b(\tan\beta=2)/\hat Y_b(\tan\beta=50)\sim1/22$ factor, 
except for the last term.

The sign of RL stop mixings is anti-correlated to $\mu$.
This can provide needed cancellations for low $m_{H^-}$, 
even if $\mu$, $M_2$ and $\widetilde m$ are large. 
For $|\mu|=100, 500$ GeV, if the signs of second and third terms of 
Eq. (\ref{chargino}) are negative, the charged Higgs mass can be as low as
300, 400 GeV,
for a rate within 20\% from SM expectation.
Even for $|\mu|=1$ TeV, the cancellation may lower $m_{H^-}$ to 600 GeV. 
The cancellation, however, requires some degree of fine-tuning.
Without such cancellations,
we would need to require $|\mu|,m_{H^-}>1$ TeV if
we allow deviations from the SM rate to be within 20\%.
  
For $b\to d\gamma$ decay, 
we should replace $V^*_{ls,ts}$ and $\lambda^2$ in Eq. (\ref{chargino}) 
by $V^*_{ld,td}$ and $-\lambda^3 e^{i\phi_1}$, respectively,
while Eq.~(\ref{chargedHiggs}) remains unchanged.
For $V^*_{ls} V_{mb} \delta^{lm}_{uLL}$ real and negative,
it would cancel against charged Higgs contribution for low $\mu$.
However, this need not be the case.
For example, for $|\mu|=100$ GeV, $m_{H^-}=400$ GeV, 
the cancellation mechanism gives $Br(B\to X_s\gamma)$ within experimental error,
while $Br(b\to d\gamma)$ is enhanced by a factor of 2.
But in both $b\to s\gamma,d\gamma$ decays, 
the chargino contribution to asymmetry
$a_{M^0\gamma}$ of Eq. (\ref{amix}) is within 2\%.
 
To obtain large asymmetry $a_{\rho\gamma}$, we need a sizable   
$\sin 2 \vartheta$ (see Eq. (\ref{s2theta})) 
which requires $C_{7\gamma}$ and $C^\prime_{7\gamma}$  
to be of comparable size.  
To achieve this the New Physics must have large $C^{\prime\,\rm New}_{7\gamma}$  
but a relatively small contribution to $C_{7\gamma}$,  
since the latter already receives a large SM contribution.
For example, in case of gluinos,  
we need large $\delta^{q3}_{dRR}$ and/or   
$\delta^{q3}_{dRL}\,m_{\tilde g}/m_b$ and  
small $\delta^{q3}_{dLL}$ and $\delta^{q3}_{dLR}\,m_{\tilde g}/m_b$.  
It is interesting that this indeed can be realized in AHS models.
For similar reasons we do not expect a large modification of
$C_{7\gamma}$ from squark mixings as was noted in our discussion of 
charged Higgs effects.

To use the formula of $a_{\rho\gamma}$ we also need to know the phase.  
The SM phase in $B\to X_d \gamma$ is rather complicated since  
$u$ and $c$ quark contributions at NLO are not CKM suppressed.  
However, as shown in \cite{Ali} the NLO contribution are found to increase  
the rate by 10 \% and the long-distance contribution from intermediate  
$u$ quarks in the penguin is expected to be small.  
For exclusive modes, some model   
estimations (such as from Light-Cone QCD sum rule) 
give long-distance effect in $B\to \rho\gamma$ and  
$B\to \omega\gamma$ at about $\cal O$(15\%) \cite{LCQCD}.    
Even though long distance physics may enter, it does not enhance 
$C^{\prime\rm SM}_{7\gamma}$ \cite{LCQCD,Grinstein,smallCprime}. 
For charge $B$ decays the dominant long distance contribution come from
weak annihilation diagrams 
(giving $|C^\prime_{7\gamma}/C^{\rm SM}_{7\gamma}|\sim 4\%$), 
which is, however, absent in the case of $B^0\to\rho^0\gamma$ decay.

\begin{figure}[t]
\centerline{\DESepsf(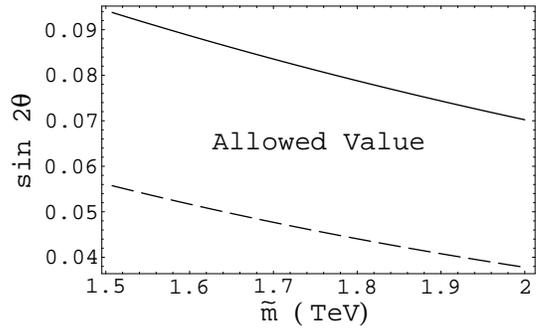 width 7cm)}
\smallskip
\caption{Asymmetry coefficient $\sin 2\vartheta$ vs. $\widetilde m$,
for gluino mass $m_{\tilde g} = 1.5$ TeV, 
where solid (dashed) curve correspond to 
$\delta^{13}_{RL}$ and $\delta^{13}_{RR}$ having opposite (same) phase.   
}
\label{fig:amix}
\end{figure}

In Fig. \ref{fig:amix}, we show gluino and neutralino contributions to
the asymmetry coefficient $\sin 2 {\vartheta}$ vs $\widetilde m$, 
with gluino mass $m_{\tilde g}=1.5$ TeV.
We use $\mu=1$ TeV and $\tan\beta=2$.
The asymmetry is generated mainly from the RL mixing induced gluino
penguins $\sim 8\%$ and can reach 10\% when including RR mixing contributions. 
The asymmetry can be measured at B Factories, and at future hadron collider
B detectors such as LHCb or BTeV.
Since $\sin2\vartheta\sim 2 |C^\prime_{7\gamma}/C_{7\gamma}|$ 
for small $\vartheta$,
we obtain $|C^\prime_{7\gamma}/C^{\rm SM}_{7\gamma}|$ of about 4\%,
for $\widetilde m\sim 1.5$ TeV, 
which is slightly larger than the estimation $\sim 2\%$ for long distance
effects from charm penguin \cite{Grinstein}.

\subsection{Non-standard C-Term and $\tan\beta$ Enhancement}

In a previous study, we obtain large or even maximal
asymmetry rather easily in $b\rightarrow s\gamma$ with sub-TeV superparticle
mass scale \cite{b2sp}. 
Here, the high SUSY scale as required by meson mixings 
leads to too severe a suppression in $1/G_F\widetilde m^2$,
as can be seen from Eqs. (\ref{c7New}), (\ref{c8New}). 
We find, however, that it is still possible to   
have large $a_{M^0\gamma}$ when considering   
{\it non-standard soft breaking terms} \cite{Hall}. 
Non-standard soft breaking terms can survive without inducing quadratic 
divergence if there is no gauge singlet particles in the low energy
spectrum. For scales below the horizontal symmetry breaking scale
and the masses of $S_i$, we are left with particles of the minimal 
supersymmetry standard model (MSSM).
Therefore, by using a low energy effective theory of SUSY, it is legitimate
to include these non-standard soft breaking terms, in particular,
a non-holomorphic trilinear term, which is called the $C$-term.

Besides a (standard) $A$-term,   
$A_d \langle H_d\rangle Y^\prime \tilde D_L \tilde D_R^*,$   
we now allow $(\widetilde M^2_d)_{LR}$ to have a non-standard $C$-term,  
$C \langle H^*_u\rangle Y^\prime \tilde D_L \tilde D_R^*.$  
It is natural that $A_d \sim C \sim \widetilde m$, hence  
$(\widetilde M^2_d)^{ij}_{LR} \sim \widetilde m\,M^{ij}_d\,\tan \beta$.
In this way, one gains a   
$\tan\beta \equiv \vert\langle H^*_u\rangle/\langle H_d\rangle\vert$  
enhancement factor, while  
$(\widetilde M^2_u)^{ij}_{LR}\sim \widetilde m\,M^{ij}_u$ is unaffected.  
$M_q$, and hence $U_{qL,R}$, is also unchanged,   
so the previous result for $D^0$ mixing remains unchanged.   
Some zeros in $(\widetilde M^2_q)_{LR}$ may also be lifted   
since these $C$-terms are no longer holomorphic,   
but they are still suppressed.  
We note that the $\delta^{12}_{dRL}$ contribution to   
kaon mixing remain protected by   
the smallness of $M^{21}_d/{\widetilde m}$.  
Likewise, for $B_d$ and $B_s$ mixings,  
$\tan\beta$ enhancement of $\delta_{dLR,RL}^{i3}$  
is insufficient to overcome $m_q/\widetilde m$ suppression  
and $\delta^{i3}_{dRR}$ still dominates.    

\begin{figure}[t!]  
\centerline{\DESepsf(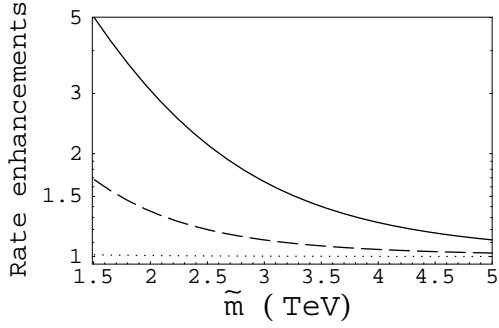 width 7cm)}  
\smallskip  
\caption{The rate enhancement of  
Br$(B\to X_d\gamma)$ with respect to SM results
with non-standard soft breaking terms.  
The solid, dashed and dotted curves
correspond to $\tan\beta=50,\,20$ and 2, respectively.   
The first curve can enhance the rate up to a factor 5,
while the second one can enhance the rate up to a factor 1.8.
The enhancement factor for the third one is below 10\%.  
}  
\label{fig:rate}
\end{figure} 

\begin{figure}[t!]  
\centerline{\DESepsf(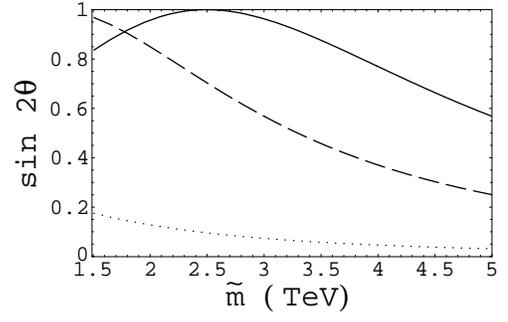 width 6.8cm)}  
\smallskip  
\caption{The CP violating coefficient $\sin 2\vartheta$ with respect   
to $\widetilde m$ for gluino and neutralino contributions.
We use $m_{\tilde g}=1.5$ TeV.
These curves use the same parameter space as the previous figure.  
Note that first two curves may reach their maximal value  
even with a multi-TeV squark mass,
while the small $\tan\beta$ case may gives 20 \% asymmetry.} 
\label{fig:s2theta}
\end{figure}

We illustrate in Figs. \ref{fig:rate} and \ref{fig:s2theta} the ratio  
${\rm Br}(B\to X_{d}\gamma)/{\rm Br}(B\to X_{d}\gamma)_{\rm SM}$  
and the coefficient $\sin 2\vartheta$ 
relevant for mixing dependent $CP$ violation,
with respect to the average squark mass $\widetilde m$ 
for $m_{\tilde g} =1.5$ TeV.  
The solid, dashed and dotted curves correspond to
$\tan\beta=50,\,20$ and 2, respectively.
The branching ratio can be enhanced by a factor of 5   
with respect to the SM value.
This can be easily understood by noting that, 
before introducing the $C$-term, 
the RL mixing induced gluino diagrams give 
$|C^\prime_{7\gamma}/C^{\rm SM}_{7\gamma}|\sim 4\%$ for
$\widetilde m\sim 1.5$ TeV.
Adding the nonstandard $C$-term 
enhances $\delta_{dRL}$ by $\tan\beta$ and hence 
$|C^\prime_{7\gamma}/C^{\rm SM}_{7\gamma}|$
is brought up to $0.04\,\tan\beta$.
A factor of 5 enhancement in rate follows for $\tan\beta=50$.
Note that $\sin 2\vartheta$ reaches maximum for
$\widetilde m\sim 2.6$ TeV. The reason is simply because
$C^\prime_{7\gamma}$ dominates over $C_{7\gamma}$ for 
lower $\widetilde m$ scale, hence suppresses the asymmetry 
while enhancing the rate significantly. 
Since the phase combination $\sin[2\Phi_B-\phi(C_7)-\phi(C^\prime_7)]$ 
in general should not vanish even if $\phi(C^{(\prime)}_7)$ vanishes 
(because of non-vanishing $\Phi_B$), 
$a_{\rho^0\gamma}$ could clearly be sizable, 
which would unequivocally indicate the presence of New Physics.

The CP violating partial rate asymmetry $A_{\rm CP}$   
in $b\rightarrow d\gamma$ decay is defined as   
\begin{eqnarray}  
A_{{\rm CP}} &=&  
{\frac{\Gamma (b\rightarrow d\gamma )-\bar{\Gamma}(\bar{b}  
\rightarrow \bar{d}\gamma )}{\Gamma (b\rightarrow d\gamma )  
+\bar{\Gamma}(\bar{b}\rightarrow \bar{d}\gamma )}}  
\nonumber \\
&=&{\frac{|C_{7\gamma}|^{2}+|C_{7\gamma}^{\prime }|^{2}  
 -|\bar{C}_{7\gamma}|^{2}-|\bar{C}_{\gamma7}^{\prime }|^{2}}{|C_{7\gamma}|^{2}  
 +|C_{\gamma7}^{\prime }|^{2}+|\bar{C}_{\gamma7}|^{2}  
 +|\bar{C}_{7\gamma}^{\prime }|^{2}}},  
\label{acp}
\end{eqnarray}  
where $\bar{C}_{7\gamma}^{(\prime )}$ are coefficients for   
$\bar{b}$ decay.  
To have nonzero $A_{{\rm CP}}$, apart from CP phases,  
one also needs absorptive parts.   
In the model under consideration, these can   
come only from the SM contribution with $u$ and $c$ quarks in the loop.  
The $A_{{\rm CP}}$ is smaller than the SM one
since $\delta_{LR}$ which contributes to  
$C_{7\gamma}$ is much smaller than $\delta_{RL}$,
while there is no strong phase in $C^\prime_{7\gamma}$ 
to contribute to direct CP violation.  
Therefore New Physics only dilutes the $A_{{\rm CP}}$ in this case
by contributing to the total rate in the denominator of Eq. (\ref{acp}).  
That is, $A_{{\rm CP}}$ is reduced by a rate enhancement factor
for large $\tan\beta$.


These figures hold also for $b\to s\gamma$ 
for the other choice of using Eq. (\ref{Mdb2sp})
with $\tilde s$--$\tilde b$ but no $\tilde d$--$\tilde b$ mixing.
Allowing for 20\% rate uncertainty for the measured Br$(B\to X_s\gamma)$,  
we see that for the $\tan\beta=20$ case,
$\widetilde m \geq 3$ TeV is allowed,  
while for heavier squark $\widetilde m=5$ TeV
the full range of $2 \lesssim\tan\beta \lesssim 50 \sim m_t/m_b$ is allowed. 
In these cases $\sin 2\vartheta$ can go up to $\sim 0.6$.
For lighter $\widetilde m$ such as 1.5 TeV,  
the enhancement factor for large $\tan\beta$ starts to  
break the good agreement between SM and   
the experimental value of Br$(B\to X_s\gamma)$,  
hence it seems one cannot have both large $\tan\beta$ and  
$\tilde m$, $m_{\tilde g}$ too light (approaching TeV).

The interesting case of ``strange-beauty" squark \cite{sb}
where large mixing dependent $CP$ 
is possible without non-standard $C$-terms
is discussed in next section.

\section{Discussion} 

We offer to discuss a few miscellaneous items.

\subsection{Re$({\varepsilon^\prime / \varepsilon})$ and EDM Constraints}

Unlike Ref. \cite{Masiero},  
the AHS model presented here can not be responsible for  
the large value of ${\rm Re}({\varepsilon^\prime / \varepsilon})$.  
This is because of heavy masses and  
the suppressed value of $\delta^{12}_{dLR,RL}$,  
as shown in Table II.  
However, ${\rm Re}({\varepsilon^\prime / \varepsilon})$ does not provide
further constraint on $\delta$'s,   
since we already satisfy the most severe case, 
$\varepsilon_K<2.268\times10^{-3}$.  
For further discussion on the issue of 
${\rm Re}({\varepsilon^\prime / \varepsilon})$  
in the context of AHS models, see Ref. \cite{eyal}.  

There are other horizontal symmetry models that lead to 
the pattern of quark mass ratios and mixings.  
In particular, the non-Abelian U(2) horizontal model \cite{U2}  
gives suppressed $\delta^{12}_{dLL,RR}$ and
can evade the kaon mixing constraint by a U(2) symmetry  
with relatively light masses.  
Mixing angles in fermion mass matrices are in general of the order of  
the square root of mass ratios \cite{U2},  
\begin{eqnarray}  
M_q&=&U^\dagger_{qL} M_{\rm diag} U_{qR},  
\\  
U_{qL,R}&=&\left(  
\matrix{1                          &s^{12}_{qL,R}                 &0 \cr  
	-s^{12}_{qL,R}              &1                 &s^{23}_{qL,R} \cr  
	s^{12}_{qL,R}s^{23}_{qL,R}   &-s^{23}_{qL,R}     &1}  
	\right),  
\end{eqnarray}  
where  
\begin{equation}  
s^{23}_{qL} s^{23}_{qR}=\left({m_2\over m_3}\right)_q,  
\quad  
-s^{12}_{qL}=s^{12}_{qR}=\sqrt{\left({m_1\over m_2}\right)_q}.  
\end{equation}  
The model leads to large Re($\varepsilon^\prime/\varepsilon$) \cite{murayama}.  
The $D$-$\bar D$ mixing is small in the same way that 
the kaon mixing constraint is evaded. 
Mixing in $\tilde b$--$\tilde d$ is relatively small,
but the sparticle scale can be relatively light 
since kaon and $D$ meson mixing constraints are evaded.
This model may also lead to FCNC effects in B system \cite{Masiero:2001cc}. 

Let us now consider the constraint from the electric dipole moment (EDM) 
of the neutron. It is well known that the EDMs of
the electron and atoms give severe constraint on SUSY phases \cite{edm}.
This is a common problem to all SUSY models and is quite independent from
FCNC processes considered in this work.
The problem should not be worse in our case,
and in fact the TeV sparticle scale should loosen
the constraint compared with usual considerations.

For the neutron EDM, there are contributions from electric dipole operator, the
color dipole operator and the dimension six purely gluonic operator. We expect
the first one to be dominant while the others may give comparable contributions
and may in some cases loosen the constraint through cancellations
\cite{Ibrahim,Weinberg}.
In addition, there are two loop contributions \cite{Chang}. 
These contributions could become important in the absence of 
large one loop contributions, such as in SUSY models with massive first
and second generation squarks \cite{Baek:1999yn}.
But since we have potentially large one loop contributions,
we use only the electric dipole operator to 
estimate the order of magnitude bounds. 
For a more complete recent study of EDM constraint on SUSY models, 
see Ref. \cite{Abel:2001vy}.

The neutron EDM can be expressed as $d_n= {1\over 3}\eta^E (4 d_d-d_u)$,
where $\eta^E=1.53$ is the QCD correction factor \cite{Ibrahim}. 
For gluino contribution, we have \cite{Gabbiani,Ibrahim},  
\begin{equation}
{d_q\over |e|}={2Q_{\tilde q}\alpha_s m_{\tilde g}\over 3\pi\tilde m^2}\,  
            \, g_4(x_{\tilde g\tilde q})\,{\rm Im}\,\delta^{11}_{qLR},
\end{equation} 
and for chargino \cite{Ibrahim},
\begin{eqnarray}  
{d_u\over |e|}&=&\sum^2_{i=1}{\alpha_w m_{\tilde \chi^+_i}\over 4\pi\tilde m^2}\,  
            \, [Q_{\tilde d}\,g_4(x_{\tilde \chi^+_i\tilde q})
\nonumber \\
&&\qquad\qquad+(Q_{\tilde d}-Q_u)\,g_3(x_{\tilde \chi^+_i\tilde q})]
             \,{\rm Im}\,\eta^{\tilde \chi^+}_{\tilde d},  
\\
\eta^{\tilde \chi^+}_{\tilde d}&=&-\hat Y_u {\cal V}^*_{j2} 
\,({\cal U}^*_{j1} V_{ul} \delta_{dLL}^{lm} V^*_{um}
   -{\cal U}^*_{j2} V_{ul} \delta^{lm}_{dLR} V^*_{um} \hat Y_{d_m}).
\nonumber \\  
\end{eqnarray} 
where $l,m$ are summed over three generations of down type squarks.
By interchanging 
$Q_{u,\tilde d}$, ${\cal U},V_{ul}\delta^{lm}_{dAB} V^*_{um}$, 
$\hat Y_{d_m}$  with 
$Q_{d,\tilde u}$, ${\cal V},V^*_{u_l d}\delta^{lm}_{uAB} V^*_{u_m d}$,
$\hat Y_{u_m}$, respectively, one can obtain $d_u$.
Finally, for neutralino contributions, one finds \cite{Ibrahim},
\begin{eqnarray}  
{d_q\over |e|}&=&\sum^4_{i=1}
         {Q_{\tilde q}\alpha_w m_{\tilde \chi^0_i}\over 4\pi\tilde m^2}\,  
            \, g_4(x_{\tilde \chi^0_i\tilde q})   
             \,{\rm Im}\,\eta^{\tilde \chi^0}_{\tilde q},  
\\
\eta^{\tilde \chi^0}_{\tilde q}&=&-2 G^{*i}_R \delta^{11}_{qRL} G^i_L  
                                  -{\sqrt2}G^{*i}_R\delta^{11}_{qRR} H^i_R
\nonumber \\                                  
&&\qquad\qquad+{\sqrt 2}H^{*i}_L \delta^{11}_{dLL} G^i_L
                                  +H^{*i}_L \delta^{11}_{qLR} H^i_R.  
\label{neutralino_edm}
\end{eqnarray}

To illustrate the constraint on squark mixing phases we 
take chargino and neutralino mixing matrices to be real, 
and discuss the phase of $\mu$ later. 
Requiring $d_n$ to be less than the current experimental bound of
$0.63\times10^{-25}\,e$ cm \cite{PDG}, we obtain  
\begin{eqnarray}  
|{\rm Im}\,\delta^{11}_{dLR}|&\leq& (2.8,\,3.4,\,6.5)\times10^{-6},  
\\
|{\rm Im}\,\delta^{11}_{uLR}|&\leq& (5.5,\,6.7,\,12.6)\times10^{-6},
\end{eqnarray}  
for $\widetilde m=1.5$ TeV, $|\mu|=100$--1000 GeV, and  
$x_{\tilde g\tilde q}=(0.3,\,1,\,4)$, respectively.
These bounds are consistent with Ref. \cite{Abel:2001vy}.
The bounds come dominantly from gluino contributions and 
hence insensitive to $\tan\beta$ and $|\mu|$.
From chargino contribution alone with above parameter space and 
$\tan\beta=50\,(2)$, we have 
\begin{eqnarray}
|{\rm Im}\,(\sum_{lm} V_{ul} \delta^{lm}_{dLR} V^*_{um} \hat Y_{d_m})|
&\leq& 0.43-0.54\;(0.39-0.48),
\nonumber\\ 
|{\rm Im}\,(\sum_{lm} V^*_{ld}\delta^{lm}_{uLR} V_{md} \hat Y_{u_m})|
&\leq& 0.11\;(0.10).
\end{eqnarray}
The AHS model gives  
$|\delta^{11}_{dLR}|\sim 
m_d|A_d(1+\{\tan\beta\})-\mu\tan\beta)|/\widetilde m^2
\sim |m_d/\widetilde m|\,\tan\beta 
\sim 8.4\times 10^{-5}(\tan\beta/50)$.			        
Thus, for large $\tan\beta$, we need $\arg(\delta^{11}_{dLR}$) to be less 
than 0.1 to satisfy the EDM constraint.  
EDM from Mercury atom gives $d_{H_g}<2.1\times 10^{-28}$ e cm \cite{Romalis:2001mg}.
The bounds on $|{\rm Im}\delta^{11}_{u,dLR}|$ are one order of magnitude smaller
than that from the neutron EDM bounds \cite{Abel:2001vy},
\begin{equation}
|{\rm Im}\,\delta^{11}_{dLR}|\leq (1.1,\,2.0,\,4.5)\times10^{-7}, 
\end{equation}
for 1.5 TeV $\msq$.
Therefore, $\arg(\delta^{11}_{dLR}$) in this model need to be 
smaller than 0.01 for large $\tan\beta$.

If $\mu$ is complex, it will contribute to $\arg(\delta^{11}_{dLR}$) as
$-\arg(\mu) \,\tan\beta\, m_d|\,\mu|/\widetilde m^2$.
For large $\tan\beta$ and $|\mu|\sim\widetilde m$,
we need arg($\mu$) to be less than 0.1 (0.01) from 
the neutron (Mercury) EDM constraint.
One should be more concerned, however, with the presence of
$\delta^{11}_{dRR(LL)}$ in Eq. (\ref{neutralino_edm}).
Take  $\delta^{11}_{dRR}$ for example,
we note that it is $\sim{\cal O}(1)$ and is 
not suppressed by quark mass like $\delta^{11}_{dLR(RL)}$.
Thus, it will lead to a severe constraint on $\arg(\mu$).
For $\mgl=\msq=1.5$ TeV, $\tan\beta=2$ and $|\mu|=100$--1000 GeV, 
we need to have arg($\mu)\leq 0.03$--$0.012$. 
The bound is roughly inversely proportional to $\tan\beta$.
For large $\tan\beta$, say 50, 
we need $\arg(\mu)\leq 5\times10^{-4}$ from the neutron EDM constraint. 
This constraint is quite severe, even for $\mgl,\msq$ at TeV scale. 
However, the very strong constraint on $\arg(\mu$) from EDM consideration is a
well known problem (see for example, Ref. \cite{Abel:2001vy}),
and is not aggravated by considerations of FCNC induced by squark
mixings, which has been the main focus of our study.

\subsection{Radiative $c\to u\gamma$ and $t\to c\gamma$ Decays}

It is of interest to check radiative flavor changing neutral current processes   
in up type quark decays, since quark-squark alignment has 
shifted flavor violation to the up-type sector.  
It is well known that the short distance one-loop $c\to u\gamma$   
amplitude is very small in the SM,   
due to the CKM suppression and the small $m^2_b/M^2_W$ factor.   
The amplitude can be raised by 2 orders of magnitude 
when one considers leading   
logarithmic QCD corrections involving operator mixings,   
and further raised by another 3 orders of magnitude when  
including non-CKM suppressed two-loop diagrams \cite{greub}.  
It is also known that long distance effects are in general large \cite{he}.      
 
We can obtain SUSY contribution by using formulas similar to $b\to q\gamma$,
\begin{eqnarray}  
H_{{\rm eff.}} &=&-{\frac{G_{F}}{\sqrt{2}}}{\frac{e}{4\pi ^{2}}}%
m_{c}\,\bar{u}\left[ c_{7\gamma}\,R+  
c_{7\gamma}^{\prime}\,  
L\right] \sigma _{\mu \nu }F^{\mu \nu }c  
\nonumber \\  
&&\hspace{-0.2cm}-{\frac{G_{F}}{\sqrt{2}}}{\frac{g}{4\pi^{2}}}m_{c}  
\,\bar{u}\left[ c_{8g}\,R+  
c_{8g}^{\prime }L\right]  
\sigma _{\mu \nu }T^{a}G_{a}^{\mu \nu }c, 
\end{eqnarray}
Note that here we do not factor out the CKM factor from 
$c^{\prime}_{7\gamma,8g}\,$s 
(hence use lower case symbol) in the Hamiltonian.
For chargino contributions we have,
\begin{eqnarray}
&&c_{7\gamma,\tilde\chi^-}=  
{m^2_{\rm W}\over \widetilde m^2}
\nonumber \\  
&&\times\biggl\{\biggl[{\cal U}_{j1}{\cal U}^*_{j1} V_{ul}\delta^{lm}_{dLL} V^*_{cm}
-{\cal U}_{j1} {\cal U}^*_{j2} V_{ul} \delta^{lm}_{dLR} V^*_{cm}\hat Y_{d_m}
\nonumber \\  
&&\,\,-{\cal U}_{j2} {\cal U}^*_{j1} V_{ul} \hat Y_{d_l} \delta^{lm}_{dRL} V^*_{cm}    
+{\cal U}_{j2} {\cal U}^*_{j2} V_{ul} \hat Y_{d_l} \delta^{lm}_{dRR} V^*_{cm}
\hat Y_{d_m}\biggr]\times
\nonumber \\  
&&\,\,
\biggr[Q_d g_2 (x_{\tilde\chi^-_j\tilde q})-g_1 (x_{\tilde\chi^-_j\tilde q})\biggr]  
-{m_{\tilde\chi^-_j}\over m_c} \biggl[{\cal U}_{j1} {\cal V}_{j2} 
V_{ul} \delta^{lm}_{dLL} \hat Y_c V^*_{cm} 
\nonumber \\
&&\,\,-{\cal U}_{j2}{\cal V}_{j2}V_{ul}\hat Y_{d_l}\delta^{lm}_{dRL}
\hat Y_c V^*_{cm}\biggr]  
\biggl[Q_d g_4 (x_{\tilde\chi^-_j\tilde q})-g_3 (x_{\tilde\chi^-_j\tilde q})\biggr]  
\biggr\},  
\nonumber \\
&&c^\prime_{7\gamma,\tilde\chi^-}=  
\hat Y_u {m^2_{\rm W}\over \widetilde m^2}
\nonumber \\  
&&\times\biggl\{{\cal V}_{j2}{\cal V}^*_{j2}V_{ul}\delta^{lm}_{dLL} \hat Y_c V^*_{cm}
\biggl[Q_d g_2 (x_{\tilde\chi^-_j\tilde q})-g_1 (x_{\tilde\chi^-_j\tilde q})\biggr]
\nonumber \\
&&\,\,
+{m_{\tilde\chi^-_j}\over m_c}
\biggl[Q_d g_4 (x_{\tilde\chi^-_j\tilde q})-g_3 (x_{\tilde\chi^-_j\tilde q})\biggr]
\\  
&&\times  
\biggl[{\cal V}^*_{j2} {\cal U}^*_{j2}    
 V_{ul} \delta^{lm}_{dLR} V^*_{cm}\hat Y_{d_m}
-{\cal V}^*_{j2}{\cal U}^*_{j1} V_{ul} \delta^{lm}_{uLL} V^*_{cm}\biggr]    
\biggr\},
\nonumber  
\end{eqnarray}
where we sum over $l,\,m$ for three generations and $j$ for
two chargino mass eigenstates. 
One can obtain $c^{(\prime)}_{8g}$ by replacing $g_{1,3}(x)$ with zero and
$Q_d$ by one.
Comparing these equations to those in the previous section,
we have interchanged ${\cal U}$ with ${\cal V}$ and modified the charge factor
in front of $g_i$'s.
In general, $c^\prime_{7\gamma,\chi^-}$ is small due to the smallness of 
$\hat Y_u$.
The chargino loop contribution on $c_{7\gamma}$ is dominated by 
LL mixing.
For $m_{\tilde g}=\widetilde m=1.5$ TeV, $|\mu|=100$--1000 GeV, 
$\tan\beta=2$--$50$ and $V_{ul} \delta^{lm}_{dLL} V^*_{cm}\sim\lambda$ 
the chargino loop gives $|c_{7\gamma}|\sim 10^{-4}$,
which is one order of magnitude below the Cabibbo favored two-loop
amplitude.
Contributions from LR and RL mixings are smaller 
by a few orders of magnitude compare to the LL mixing contribution.
Therefore, the $\tan\beta$ enhancement effects are small and unable to
overcome the heavy superparticle decoupling effects.
This is also true in other up-type FCNC processes,
such as $t\to c\gamma$ to be discussed later.
Charged Higgs contribution is small since we do not have the top quark
in the loop. 

Formulas for gluino and neutralino contributions are 
similar to previous section with a trivial modification on neutralino
mixing matrices $G_{L,R}$ and $H_{L,R}$. 
For $\widetilde m=m_{\tilde g}=1.5$ TeV, 
gluino and neutralino loops give $|c_{7\gamma}|\sim 10^{-6}$,
which contribute to Br($c\rightarrow u\gamma$) at
the same order of magnitude of the leading logarithmic SM result.

For the $t\to c\gamma$ case, the SM result is very small,   
Br($t\to c\gamma)\sim 10^{-13}$ \cite{eilam}.  
In this model, by using similar formulas for $c^{(\prime)}_{7\gamma}$, 
we have,
\begin{equation}
\Gamma(t\to c\gamma)={G_F^2\alpha\over 32\pi^4} m_t^5
(|c_{7\gamma}|^2+|c^\prime_{7\gamma}|^2).
\end{equation}
For the parameter space considered in the previous case, 
the chargino loop gives Br($t\to c\gamma)\sim 10^{-9}$, dominated by 
chiral enhanced LL mixing.
Therefore, it is not sensitive to the non-standard soft breaking term.
The rate is still unobservable.
The gluino contribution is as small as the SM one.  
This is in contrast to generic MSSM with non-universal soft squark masses
where the rate can be close to experimental bounds \cite{deDivitiis:1997sh}.

We see that, even though the flavor violation is 
shifted to the up-type sector, 
we still do not have large FCNC nor CP violation in $t$, $c$ decays.
This is because of heavy $\widetilde m$, $m_{\tilde g}$ masses 
as required by meson mixings, and 
absence of enhancement mechanisms in gluino and neutralino diagrams.

\subsection{Light Strange-Beauty Squark?}

In this paper we have focused on the general case of 
naturally large $\tilde d_R$-$\tilde b_R$ or $\tilde s_R$-$\tilde b_R$
mixings as a consequence of Abelian flavor symmetry with SUSY.
As we have seen, kaon FCNC constraints require texture zeros to 
remove $\tilde d$-$\tilde s$ mixing.
This is done by QSA, which shifts the source of Cabibbo angle
to up-type sector.
It then follows that both the $K^0$ mixing constraint and $D^0$ mixing bound
demand TeV scale SUSY particles.
In the case of $\tilde d_R$-$\tilde b_R$ mixing
(mutually exclusive with $\tilde s_R$-$\tilde b_R$ mixing),
$B_d$ mixing constraint also implies TeV scale sparticle masses,
and one could get maximal $\sin2\phi_1$
as suggested by recent Belle result \cite{Belle2}.

The $\tilde s_R$-$\tilde b_R$ mixing case has several special features
worthy of note.
First, it is maximal, largely because $V_{cb} \sim m_s/m_b \sim \lambda^2$.
Second, 
unlike $\Delta m_{B_d}$ which is precisely measured already,
we only have a lower bound on $\Delta m_{B_s}$.
In fact, data hints at $\Delta m_{B_s} > \Delta m_{B_s}^{\rm SM}$.
From the latter, one cannot draw the conclusion that
$B_s$ mixing data demand TeV scale sparticles.
From the former, it is intriguing that,
in fact, one has a mechanism for the possibility of {\it one} light squark.
As pointed out in Ref. \cite{sb}, the ``democratic" nature of 
the 2-3 sub-matrix of $\widetilde M_{dRR}$ in Eq. (\ref{MRR}) 
not only induces maximal $\tilde s_R$--$\tilde b_R$ mixing, 
it could also drive one mass eigenstate $\widetilde sb_1$,
dubbed the ``strange-beauty" \cite{sb} squark because it carries both flavors
equally, to be much lighter by level splitting.
What is rather surprising is that, having $\widetilde sb_1$
as light as 100 GeV does not make visible impact on the $b\to s\gamma$ rate.
Thus, the light $\widetilde sb_1$ scenario survives one of the
strongest known constraints on new physics!
This has phenomenological bearings.

The general average squark mass scale $\widetilde m$ is 
still fixed by $K^0$ and $D^0$ mixings at TeV.
But with some tuning in the $\widetilde M_{dRR}$ matrix,
for example $\widetilde m_{23}^2/\widetilde m^2 \sim 1$ to $\lambda^3$ order, 
$\widetilde sb_1$ can be brought down to 100 GeV.
With such large squark mass splittings,
the formulas in previous sections do not apply, 
but one can still follow Ref. \cite{Bertolini}.
In fact, we find that the Br$(b\to s\gamma)$ constraint itself
can be easily satisfied even if $m_{\widetilde sb_1}\to 0$.
We note three major consequences of experimental interest:
i) Because of low $\widetilde sb_1$ mass, sizable $C^\prime_7$ is generated.
Although it is sub-dominant in $b\to s\gamma$ rate,
it allows the mixing dependent $CP$ asymmetry, 
e.g. in $B_s\to \phi\gamma$, to go up to 60\%.
There is no need to resort to nonstandard $C$-terms in this case.
ii) A light $\widetilde sb_1$ squark further enriches
$B_s$ mixing and its associated $CP$ phase.
$\Delta m_{B_s}$ could be close to or larger than the SM expectation,
and a non-vanishing $CP$ phase in $B_s$ mixing 
can be measured readily in the moderate $x_{B_s}$ case.
iii)~A light $\widetilde sb_1$ squark clearly offers itself 
for direct search at future colliders,
{\it in a model where sparticles are otherwise at TeV scale}.
In fact, one neutralino, the bino, could also be rather light.
A possible decay hence search scenario is
$\widetilde sb_1 \to (b,s) + \widetilde \chi^0_1$
with equal probability of $s$ and $b$ quarks in decay final state.
Note that other predictions, such as sizable $x_D$ from SUSY, still hold.  

This special scenario, perhaps a bit tuned, seems tailor made
for spectacular measurements at the Tevatron and the future LHC.
More details and discussions can be found in Ref. \cite{sb}.

\section{Conclusion}

In this work,
we make a complete one-loop analysis in SUSY AHS models on FCNC
concerning $B_d$, $B_s$, $K^0$, $D^0$ mixings and 
$b\rightarrow d\gamma,\,s\gamma$ decays.   
We find that $B_d$ ($B_s$) and $D^0$   
mixings all receive sizable SUSY contributions
even with TeV scale superparticles.   

Large off-diagonal elements involving the third generation 
in the fermion mass matrices follow naturally in AHS models.  
Hence, flavor mixings involving  
$\tilde d_{jR}$ are naturally prominent.
It could be the source for near maximal
$\sin 2\phi_1$ given by recent experiments.  
For $m_{\tilde q}$ and $m_{\tilde g}$ at TeV scale,  
the effects could be comparable to SM   
in $B_d$ (or $B_s$) mixing,  
leading to $\phi_{B_d} \neq \phi_1$ (or $\phi_{B_s} \neq 0$),  
while $K^0$ mixing and $\varepsilon_K$  
require quark-squark alignment to   
make $M_d^{12}$ and $M_d^{21}$ vanish.   
This shifts $V_{us}$ to $u$ sector,  
and $\tilde u_L$-$\tilde c_L$ mixing with masses $\sim$~TeV  
gives $D^0$ mixing that is tantalizing close to  
recent hints from data.
With the same squark mixings, the chargino induced contributions 
to the kaon mixing also points to a TeV scale for superparticle masses, 
independent of $\Delta m_{B_d}$ considerations.  
There is a special variant where a ``strange-beauty" squark
is driven light by maximal $\tilde s_R$-$\tilde b_R$ mixing,
which can give rise to even more astounding phenomena,
but with little impact on $B_d$ system.
Otherwise, TeV is in general the preferred sparticle scale in this model.

With such heavy gluino and squarks,  
one has few other low energy phenomena,  
and prospects for direct production are depressing.
It is possible to have $\sim 10\%$ mixing-dependent asymmetry, $a_{M^0\gamma}$
in $b\to s\gamma$ and $d\gamma$ transitions.
In addition, these asymmetries 
are sensitive to nonstandard soft breaking terms   
via $\tan\beta$ enhancement, and asymmetries could be up to 60\% in
$b\to s\gamma$ when the Br($B\to X_s\gamma$) constraint
is taken into account.  
If we insist on non-vanishing $M_d^{31}$, 
the kaon mixing constraint requires, indirectly, that 
$s$ flavor is almost decoupled from the other down flavors.  
It is then possible that one has SUSY effects in   
$B_d$, $D^0$ but not $B_s$ mixings,  
while $a_{\rho^0\gamma}$ and $a_{\omega^0\gamma}$ could be maximal  
with rate enhancements up to a factor of five.
Alternatively,
effects could concentrate in $B_s$ system and $b\to s\gamma$
(plus $D^0$ mixing).
The phenomenology outlined here can be tested at B factories  
and the Tevatron in the next few years even if the New Physics
scale is so high such that direct searches show no effect.

\vskip 0.3cm  
\noindent{\bf Acknowledgement}.\ \  
This work is supported in part by  
the National Science Council of R.O.C.  
under Grants NSC-89-2112-M-002-063  
and NSC-89-2811-M-002-0086,  
the MOE CosPA Project, 
and the BCP Topical Program of NCTS.
CKC would like to thank the KEK theory group
for hospitality.


\end{document}